\documentclass[pra,twocolumn,showpacs,superscriptaddress,floatfix]{revtex4}

\usepackage{amsmath}
\usepackage{color}
\usepackage{times}
\usepackage{txfonts}
\usepackage{graphicx}

\mathchardef\ogon="012C%
\newcommand{\as}{a\kern-0.22em\lower.40ex\hbox{$_{\ogon}$}}

\begin{document}

\title{Ramsey interferometry with atoms and molecules: two-body 
  versus many-body phenomena}
\author{{Krzysztof G{\'o}ral}}
\affiliation{Clarendon Laboratory, Department of Physics, 
University of Oxford, Parks Road, Oxford, OX1 3PU, United Kingdom}
\affiliation{Center for Theoretical Physics, Polish Academy of 
Sciences, Aleja Lotnik\'ow 32/46, 02-668 Warsaw, Poland}
\author{{Thorsten K\"{o}hler}} 
\affiliation{Clarendon Laboratory, Department of Physics, 
University of Oxford, Parks Road, Oxford, OX1 3PU, United Kingdom}
\author{{Keith Burnett}} 
\affiliation{Clarendon Laboratory, Department of Physics, 
University of Oxford, Parks Road, Oxford, OX1 3PU, United Kingdom}

\begin{abstract}
We discuss the frequency and visibility of atom-molecule Ramsey fringes 
observed in recent experiments by Claussen {\em et al.} 
[Phys.~Rev.~A \textbf{67}, 060701 (2003)]. In these experiments 
a $^{85}$Rb Bose-Einstein condensate was exposed to a sequence of magnetic 
field pulses on the high field side of the 155 G Feshbach resonance. The 
observed oscillation frequencies largely agree with the theoretically 
predicted magnetic field dependence of the binding energy of the highest 
excited diatomic vibrational state, except for a small region very close to 
the singularity of the scattering length. Our analytic treatment of the 
experiment, as well as our dynamical simulations, follow the magnitude of the 
measured oscillation frequencies as well as the visibilities of the Ramsey 
fringes. We show that significant deviations from a purely binary dynamics, 
with an associated binding frequency, occur when the spatial extent of the 
molecular wave function becomes comparable with the mean distance between the 
atoms in the dilute gas. The experiments thus clearly identify the conditions 
under which diatomic molecules may be identified as a separate entity of the 
gas or, conversely, when the concept of binary physics in a many-body 
environment is bound to break down.
\end{abstract}
\date{\today}
\maketitle

%intro:
%interferometry is sensitive so a lot can depend on details of
%interactions and external perturbations - can we infer a lot?
%this method is dynamical so can attempt to mimic the experiment

%points:
%1.NMNLSE gets the shift
%2.the shift is always present -> extracting two-body quantities in a gas
%is dodgy
%3.NMNLSE also gets damping and decoherence
%4.Ramsey not Rabi -> exchange of particles only with the burst fraction
%5.no point in talking about oscillations at high densities due to an
%almost complete depletion of the condensate into the burst before the
%evolution period
%6. where does the linear decay come from? - dipolar relaxation of
%atoms, spontaneous decay of molecules?

%two-body: 2-channel Hamiltonian -> bound states
%many-body

\section{Introduction}

The study of ultra-cold molecules produced in atomic Bose and Fermi
gases is one of the most important developments in cold atom physics 
\cite{Levi00}. 
%In the environment of two component
%fermionic gases the production of weakly bound molecules is closely
%related to the build-up of strong correlations between unbound pairs
%of atoms in the so called BCS-BEC cross-over regime,
%a subject of numerous recent investigations
%\cite{Greiner03,JochimScience03,Zwierlein03,Regal04-2,Zwierlein04,Bartenstein04%,Bourdel04}.
One of the most successful experimental techniques to produce molecules
in atomic Bose-Einstein condensates utilised adiabatic sweeps 
of a magnetic field tunable Feshbach resonance level across the zero energy 
threshold of the colliding atoms, during which a highly excited diatomic 
vibrational bound state could be efficiently populated
\cite{Herbig03,Duerr03,Xu03,Mukaiyama03,Duerr04}. 
In this paper we study a coherent superposition of an atomic condensate and 
molecules. The production of such a coherent superposition in a degenerate gas 
of bosons, demonstrated in Refs.~\cite{Donley02,Claussen03}, is a remarkable 
achievement, given the marked difference in the relevant energy and time 
scales for condensates and molecules. These experiments employed a sequence 
of two magnetic field pulses each of which rapidly approached the position of 
the zero energy resonance (the magnetic field strength at which the scattering 
length has a singularity) on the side supporting the highly excited diatomic 
vibrational bound state. The two pulses were separated by an evolution period 
of variable duration as a function of which the oscillations in the final 
condensate populations were observed. This achievement offers the possibility 
of precise interferometric measurements of the energies of the relevant states 
through the measurement of bulk properties of the gas. It is the theory of 
such measurements we present in this paper.

We set up the theory in a way that reflects properly the analogy
with Ramsey -- separated pulse -- interferometers with the splitting
and recombination occurring during the magnetic field pulses at
the beginning and the end of the magnetic field sequence,
respectively. We give a complementary account of the experiment
which emphasises the interferometer picture and will, we believe,
be of greater utility to those analysing experiments than just the
simulation on its own can do.

To provide a simple intuitive physical description of
the experiments, we first consider the evolution of a single
pair of cold trapped atoms subjected to the experimental pulse sequence.
We show that the first pulse splits the initial state of the pair
into three different components (interferometer arms). The bound state 
component evolves most quickly. The other components are a superposition of  
excited quasi continuum states and the lowest energetic quasi continuum state,
which are the two-body analogues of the experimentally observed 
\cite{Donley02} burst of atoms and the remnant Bose-Einstein condensate, 
respectively. 
%The free components evolve much more slowly than the molecular bound state. 
After a variable period of time, referred to as the evolution period, 
during which the bound and free states evolve almost independently, the 
second pulse recombines the components. 
%We should emphasise that it is only
%during the pulses that the components are strongly coupled -- during
%the evolution period between the pulses no significant population
%exchange between the bound and unbound fractions occurs. 
The final state of the pair depends
sensitively on the phase differences built up between the amplitudes of 
the various components during the evolution period. As a consequence, 
oscillations in the resulting populations of the three components of the 
pair wave function are observed as the time between the pulses is
varied. We shall show that the oscillation frequency is determined by
the energy of the molecular bound state populated in the process.
The visibility of these oscillations, i.e.~Ramsey fringes, depends on both
the efficiency of populating the bound state during the first pulse
and of dissociating the molecules during the second
pulse of the sequence. We observe that these quantities are very
sensitive to both the details of the underlying two-body physics
and the precise duration and shape of the pulses. We provide
analytic estimates of the molecular production and dissociation
efficiencies and give an analogy between the function of the pulses
and the concept of two colour photo-association of molecules.

We extend this description of a single pair of atoms in order to apply it to 
a gas of atoms by using a coarse graining procedure. In this approach the 
populations of the different components of the gas are obtained, following 
classical probability theory, by averaging over the microscopic
transition probabilities for all pairs. We show that this 
straightforward but systematic extension of the two-body physics recovers the 
experimental observations rather well, in terms of the oscillation 
frequency, the amplitude and the relative populations of all three 
components present in the gas.

In order to disentangle the two-body and many-body aspects
of Ramsey interferometry with cold gases, we have also used the genuinely
many-body microscopic quantum dynamics approach to the dynamics of
partly condensed Bose gases \cite{Koehler02,KGB03}. We base our full 
simulations on the non-Markovian non-linear Schr\"odinger equation of the 
first order microscopic quantum dynamics approach. This equation describes 
the evolution of the condensate along with the dynamics of pair
correlations. The method describes fully and quantitatively the
evolution of the states produced by magnetic pulse shape induced
rapid changes in the Hamiltonian describing the inter-atomic
potential. In particular, the microscopic quantum dynamics
approach incorporates the entire binary collision dynamics in a 
non-perturbative manner.

We shall see that the many-body phenomena affect both the frequency and 
visibility of the Ramsey fringes. In the first step in this direction, we 
have found an analytical solution of the non-Markovian equation using a 
perturbative approximation. 
In agreement with the experimental observations, we show that the frequencies 
of the Ramsey fringes are shifted upwards with respect to the value associated 
with the energy of the weakly bound molecular state populated in the process. 
This shift persists for all values of the magnetic field, it does,
however, become hard to discern above $B_\mathrm{evolve}$ (the
value of the magnetic field strength between the pulses) higher
than roughly $158$ G. Away from the zero energy resonance the shift is
given, as one might reasonably expect, by roughly twice the
chemical potential. The shift is typically a small correction to the
oscillation frequency, and a time scale much longer than the typical pulse 
durations of the experiments \cite{Claussen03} would be required for it to be 
fully resolved. The visibility of the Ramsey fringes is reduced in comparison 
with the two-body result due to the non-linear dynamics of the atomic 
condensate mean field.

Our analytic studies apply when the experiments \cite{Donley02,Claussen03}
are operated in what we shall term the ``interferometer mode'', i.e.~when 
the magnetic field strength $B_\mathrm{evolve}$ during the evolution period 
is sufficiently far away from the zero energy resonance for 
two- and many-body phenomena to be disentangled. In this regime of 
magnetic field strengths $B_\mathrm{evolve}$ the gas is weakly interacting 
during the evolution period and no significant population
exchange between the bound and unbound components occurs. The different
components are then orthogonal as they should be in an ideal 
Ramsey interferometer. Claussen {\em et al}.~\cite{Claussen03} also 
performed measurements in the close vicinity of the zero energy resonance,
where the gas is subject to a non-equilibrium dynamics even between 
the pulses. For all experimentally studied cases we have performed full
simulations of the non-Markovian equation for the trap geometry
and atom number employed in the experiments. 
%The full simulation
%method has the full binary Hamiltonian embedded in the non-linear
%Schr\"odinger equation, and the back action onto the condensate,
%from which atoms are driven to bound and free states during the
%rapid variations in the magnetic field, is also included. 
%We pay particular attention to the cases where the magnetic field during
%the evolution period remains very close to the zero energy resonance.
In the regime very close to the zero energy resonance the non-linear aspects 
of the dynamics are especially significant, and the description of the 
experiment in terms of an ideal Ramsey interferometer is not as clear, because
the three components of the gas are no longer orthogonal. The full numerical 
simulation does not rely on this
assumption and is, therefore, essential in the comparison with
experimental data for this regime. We should emphasise that in
this region the idea that we can clearly separate molecules and
atoms is also breaking down as the spatial extent of the molecular bound state
wave function approaches the mean distance between the atoms in the gas 
\cite{TKTGPSJKB03}. This makes the simple picture invalid as well as any 
theory that relies on such a distinction. 

One might be misled in this analysis if one focused 
purely on the component of the molecular bound state in the bare 
resonance level that remains, of course, very small in extent. 
Our analysis shows, for instance, that the frequently used two level mean 
field approach of 
Refs.~\cite{Drummond98,Timmermans99,Yurovsky00,Salgueiro01,Adhikari01,Cusack02,Abdullaev03}, fails entirely
to describe the interferometric experiments of 
Refs.~\cite{Donley02,Claussen03}, despite the fact that it can provide 
accurate predictions on the final molecular fractions in Feshbach resonance 
crossing experiments \cite{KGG03,GKGTJ03}. In Ref.~\cite{Duine03PRL}  
a modified two level mean field approach was applied to the
observations of Ref.~\cite{Claussen03}, in which the energy of the bare 
Feshbach resonance level was tuned to recover the binding energy of the 
molecules produced in the experiments \cite{Donley02,Claussen03}. This 
approach recovered the magnitude of the fringe frequencies of 
Ref.~\cite{Claussen03} after adjusting the densities used in the theory. 
We believe, however, that such a treatment is 
inconsistent because any two level approach crucially lacks the excited energy 
levels of the background scattering continuum \cite{GKGTJ03}. This treatment 
obviously leads to the absence of the experimentally observed burst component 
of atoms \cite{Donley02} in the theory. The two level approach of 
Ref.~\cite{Duine03PRL} also inevitably misrepresents the 
molecular bound state \cite{footnote:Duine} produced in the experiments 
\cite{Claussen03}, as this bound state is even dominated by its component in 
the background scattering continuum.

%The actual molecular bound state is extremely large and mostly, of necessity, 
%in the open channel. An
%approach that focuses on the diabatic basis states that make up
%the resonance scattering, and not the full physical states, may
%produce a false perception of coupling between condensate and
%bound molecules for all the cases looked at in the experiment 
%\cite{Duine03PRL,Duine03}.

The Ramsey fringes predicted by our dynamical simulations of the
non-Markovian equation recover all the systematic trends observed
in the experiment. For magnetic fields away from the zero energy resonance,
the Ramsey fringes are essentially undamped on the comparatively short 
time scales of the pulse sequences and their frequencies are
only slightly shifted with respect to the value determined from
the energy of the weakly bound molecular state. Closer to the zero energy 
resonance, a significant decoherence of the Ramsey fringes has been observed 
in the experiment, a feature also predicted by our many-body approach. We 
also recover the substantial upward 
shift of the fringe frequencies for the magnetic field strengths closest to 
the zero energy resonance, for which the shift was attributed in  
Ref.~\cite{Claussen03,footnote}. We show that this shift can be 
interpreted in terms of a measure of the extent to which diatomic molecules 
in the highly excited vibrational state can be considered as a separate entity 
of the gas. Our studies also strongly indicate that the significant 
experimentally \cite{Claussen03} observed loss of condensate atoms, which is 
particularly clear for longer pulse sequences with the magnetic 
field strength $B_\mathrm{evolve}$ in the close vicinity of the zero energy 
resonance, is essentially determined by the same parameters of the binary
interactions that determine the frequency and visibility of the fringe 
pattern. This observation implies that deeply inelastic
loss phenomena such as resonance enhanced spin relaxation
of $^{85}$Rb atoms \cite{Roberts00}, are unlikely 
to play a significant role in the experiments \cite{Donley02,Claussen03}.

Both the two-body and many-body pictures presented in this paper
reveal the remarkable sensitivity of these experiments 
%this Ramsey interferometry in
%Bose-Einstein condensates 
to the details of the underlying two-body physics, 
details that are clouded over in experiments involving molecular production 
using linear crossing of a Feshbach resonance \cite{GKGTJ03,KGG03}. 
The interferometric technique, therefore, offers a rather unique probe 
of the microscopic collision physics through observing the macroscopic bulk 
properties of the gas.

%One may also miss breakdown of the approaches that focus
%implicitly on binary physics.

%The conditions for using the
%interferometer interpretation is, therefore, discussed in some detail
%along with an approximate solution of the associated theory. The extent
%to which a purely binary approach - as opposed to binary physics in the
%presence of a condensate - is also addressed. As emphasised above, the
%full equations have both these aspects included in them and we are able
%to compare their predictions. We are also, in this way, able to
%discuss the many-body aspects of the problem.

%In the period between the pulses
%the components evolve independently when the experiment is
%operating in its interferometer mode as we shall explain below.

\section{Physical origin of the Ramsey fringes}
\label{sec:PhysicaloriginoftheRamseyfringes}
In this section we provide a physical explanation for the experimental fringe 
patterns reported in Refs.~\cite{Donley02,Claussen03} in terms 
of general properties of a Ramsey interferometer. We then provide a very 
direct determination of the frequency and visibility of the Ramsey fringes 
in terms of the dynamics of a single pair of atoms. We then present the first 
order many-body corrections to the frequency and visibility obtained from the 
microscopic quantum dynamics approach of Refs.~\cite{Koehler02,KGB03}.
%, whose detailed derivations are given 
%in Appendix \ref{app:Manybodydynamics}. 
This allows us to distinguish pure two-body contributions to the 
interference between atoms and molecules from many-body phenomena related to 
the coherent nature of the partially condensed Bose gas.

\subsection{Atom-molecule coherence in a Bose-Einstein condensate}
\begin{figure}[htb]
  \begin{center}
    \includegraphics[width=\columnwidth,clip]{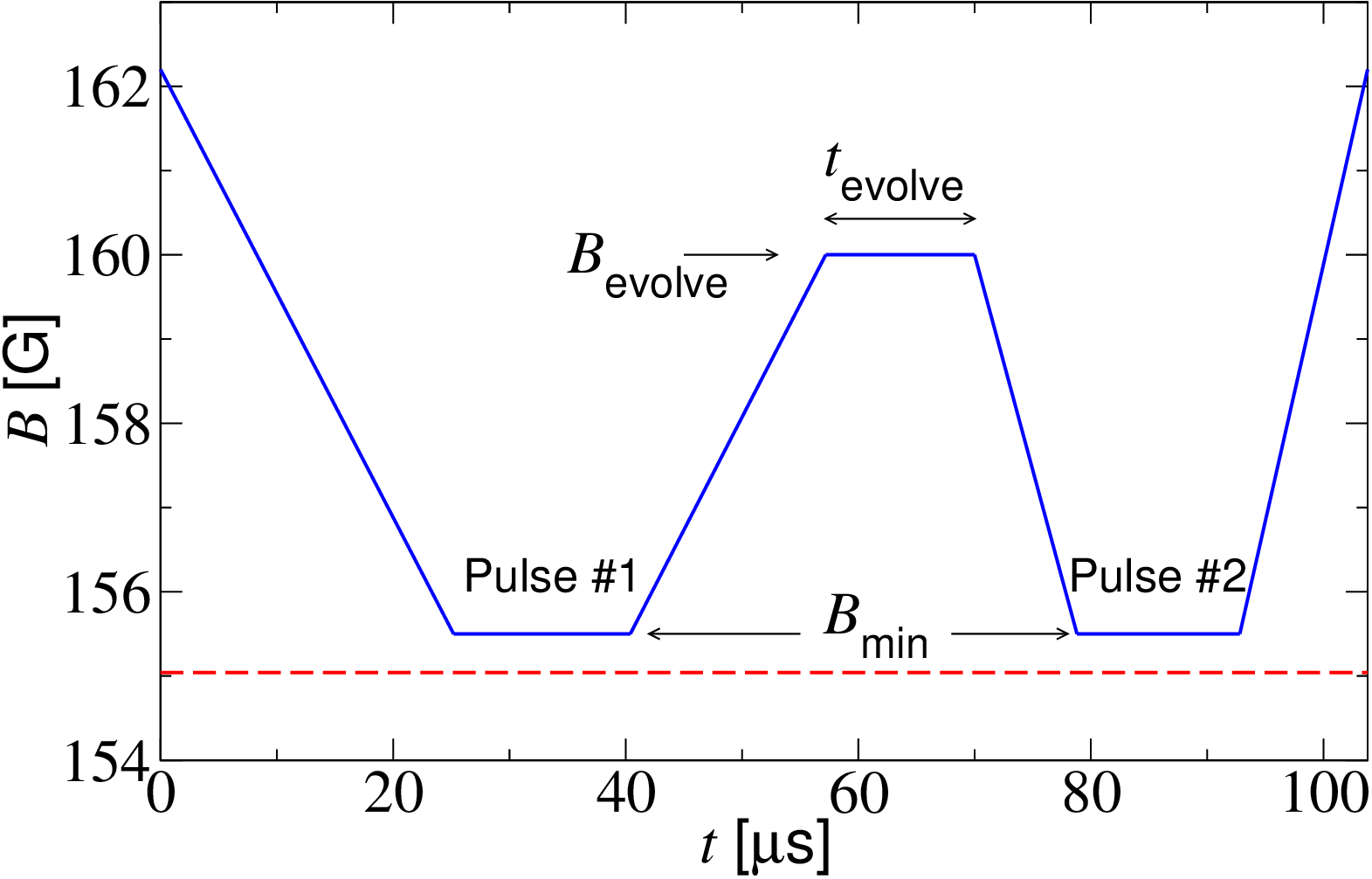}
    \caption{A typical sequence of magnetic field pulses (solid lines) applied 
      in the interferometric experiments of Refs.~\cite{Donley02,Claussen03}.
      The magnetic field strength at the minimum
      of the pulses $\#1$ and $\#2$ of $B_\mathrm{min}=155.5$ G is 
      closest to the zero energy resonance at $B_0=155.041$ G 
      \cite{Claussen03} (dashed, horizontal line), and leads to large positive 
      scattering lengths on the order of $10000\, a_\mathrm{Bohr}$
      ($a_\mathrm{Bohr}=0.052918$ nm is the Bohr radius). The 
      variable magnetic field strength and duration of the evolution period 
      of the pulse sequence are denoted by $B_\mathrm{evolve}$ and 
      $t_\mathrm{evolve}$, respectively.}
    \label{fig:pulse}
  \end{center}
\end{figure}

%\subsubsection{Interference between different components of a partially
%  condensed Bose gas}
In the experiments of Refs.~\cite{Donley02,Claussen03} a 
Bose-Einstein condensate of $^{85}$Rb atoms in the $(F=2,m_F=-2)$ hyperfine 
state was exposed to a sequence of spatially homogeneous magnetic field 
pulses on the high field side of the 155 G Feshbach resonance. 
A typical pulse sequence \cite{Hodby} is depicted in Fig.~\ref{fig:pulse}. 
The pulses, at the beginning ($\#1$) and the end ($\#2$) of the sequence, 
each consisted of linear downward and then upward ramps, which were separated 
by a hold period of constant magnetic field strength, close to the position 
of the zero energy resonance at $B_0=155.041\, \mathrm{G}$ 
\cite{Claussen03}. 
The evolution period of constant magnetic field strength $B_\mathrm{evolve}$ 
separated the pulses by an amount of time $t_\mathrm{evolve}$. 
In the course of the experiments \cite{Donley02,Claussen03}, the final 
densities of atoms were recorded after expanding the cloud, and the 
measurements were repeated with variable evolution times $t_\mathrm{evolve}$ 
and magnetic field strengths $B_\mathrm{evolve}$. 

\subsubsection{Interference between different components of a partially
  condensed Bose gas}
Donley {\em et al.}~\cite{Donley02} have identified two different components
of the final atomic cloud: a remnant Bose-Einstein condensate and a burst
of atoms with a comparatively high mean kinetic energy of roughly 
$E_\mathrm{kin}/k_\mathrm{B}=150$ nK ($k_\mathrm{B}=1.3806505\times 10^{-23}\, 
\mathrm{J}/\mathrm{K}$ is the Boltzmann constant). The magnitudes of the burst 
and remnant condensate fractions exhibited an oscillatory dependence on the 
evolution time $t_\mathrm{evolve}$ of the magnetic field pulse sequence.
The existence of a third component of missing atoms was inferred from the 
difference between the initial and total final numbers of detectable atoms.
Donley {\em et al.}~\cite{Donley02} suggested that the missing 
atoms indicated molecular production, and the oscillations were interpreted 
in terms of an interference between atoms and molecules during the evolution 
period of the pulse sequence. 

\begin{figure}[htb]
  \begin{center}
    \includegraphics[width=\columnwidth,clip]{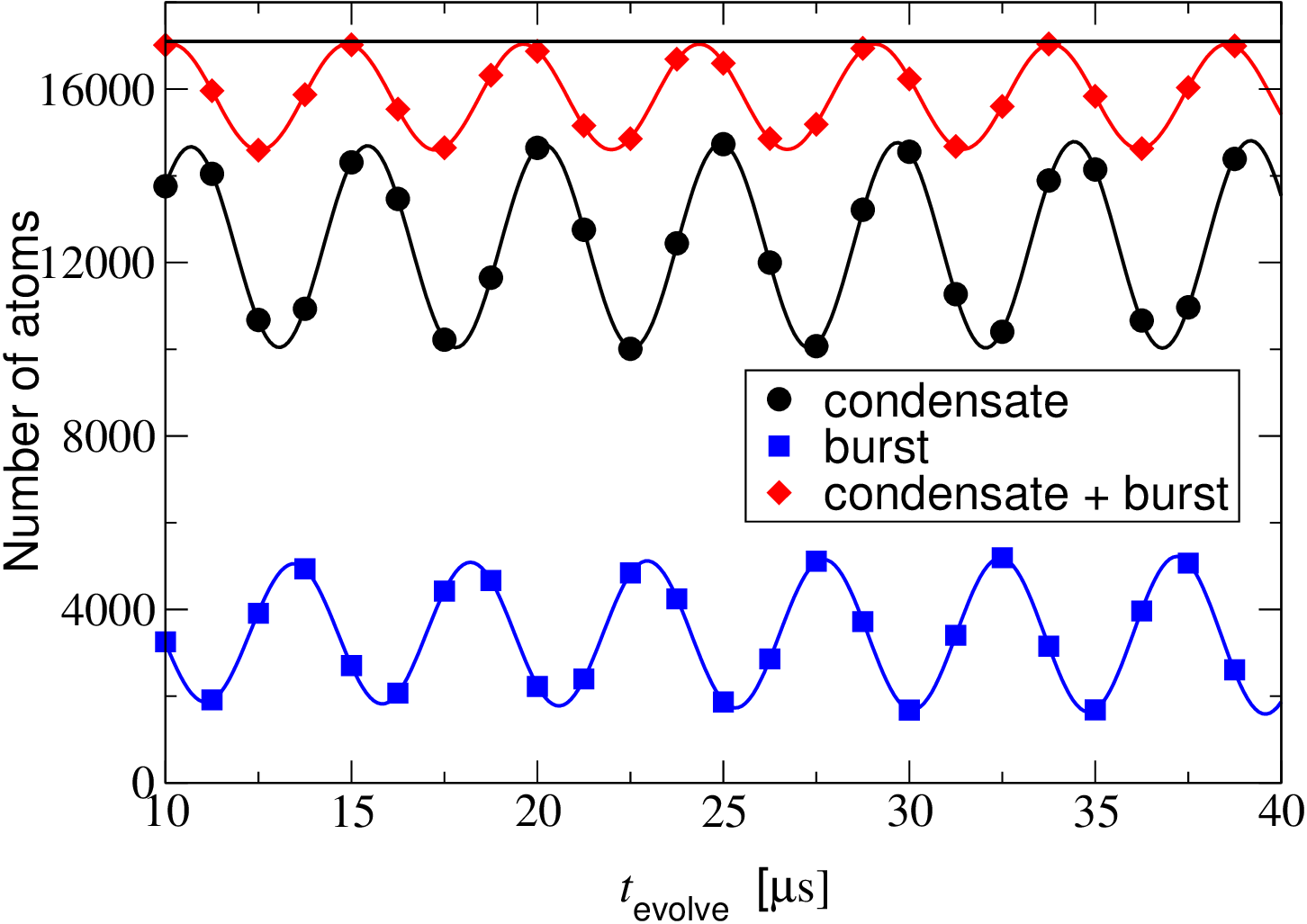}
    \caption{Predicted numbers of condensate and burst atoms at 
      the end of the magnetic field pulse sequence as a function of 
      $t_\mathrm{evolve}$ for a magnetic field strength of 
      $B_\mathrm{evolve}=159.84$ G in the evolution period of the pulse 
      sequence. The total number of atoms is 17100 (horizontal line). 
      The predictions were obtained from numerical solutions of the non 
      Markovian non-linear Schr\"odinger equation (\ref{NMNLSE}), including a 
      fully quantum mechanical description of the inhomogeneity of the gas 
      \cite{revisionKGB03} and a two-channel description of the low energy 
      binary collision dynamics. The magnetic field pulse sequence 
      (cf.~Fig.~\ref{fig:pulse}), trap frequencies, and initial ground state 
      condensate densities correspond to the measurements in Fig.~6 of 
      Ref.~\cite{Donley02}.} 
    \label{fig:85RbAMcoherence}
  \end{center}
\end{figure}

This interpretation has been confirmed by subsequent 
theoretical studies \cite{Kokkelmans02,Mackie02,KGB03}. 
Figure \ref{fig:85RbAMcoherence} shows a theoretical prediction of the 
final densities of atoms versus $t_\mathrm{evolve}$ for the pulse 
sequences and the initial conditions corresponding to Fig.~6 of 
Ref.~\cite{Donley02}. The predictions 
in Fig.~\ref{fig:85RbAMcoherence} were obtained from the microscopic quantum 
dynamics approach to dilute Bose-Einstein condensates  
\cite{KGB03,revisionKGB03}. This gives a fringe pattern which
compares rather favourably in both its qualitative and quantitative 
features with the observations reported in Ref.~\cite{Donley02}. The
same theoretical approach also recovers the experimentally observed kinetic energy
of the burst atoms on the order of $k_\mathrm{B}\times$150 nK \cite{KGB03}.
%A similar agreement between theory and experiment was achieved in 
%Ref.~\cite{Kokkelmans02}.

\subsubsection{The different regimes of molecular vibrational frequencies}
\begin{figure}[htb]
  \begin{center}
    \includegraphics[width=\columnwidth,clip]{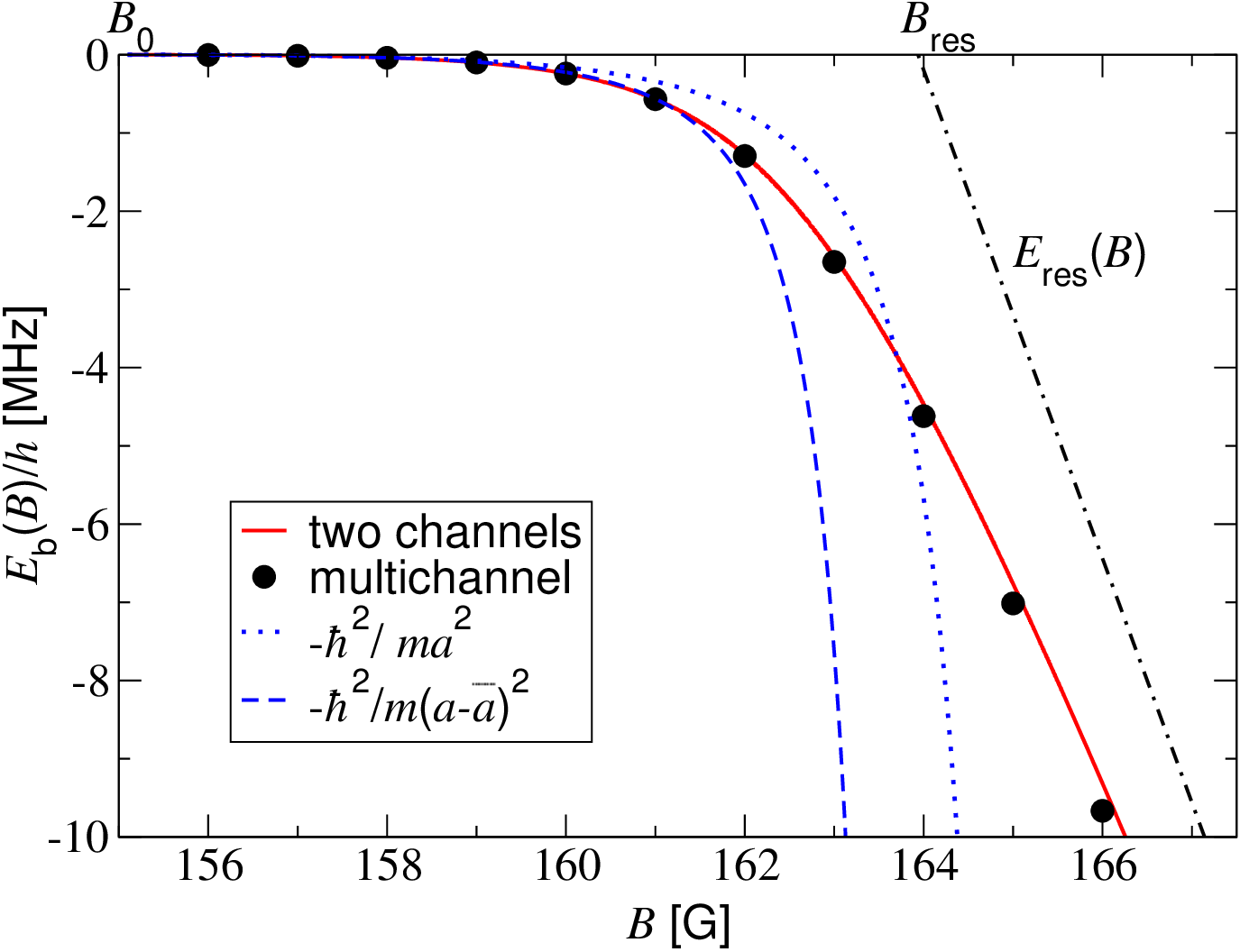}
    \caption{Binding energy $E_\mathrm{b}(B)$ of the highest exited  
      vibrational molecular bound state $\phi_\mathrm{b}(B)$ of two
      $^{85}$Rb atoms in dependence on the magnetic field strength $B$. The 
      figure shows comparisons between full coupled channels calculations 
      \cite{Claussen03,Kokkelmans} (circles), the two channel approach of 
      Refs.~\cite{KGG03,GKGTJ03} (solid curve), and single channel analytic 
      estimates \cite{KGB03} of the binding energy (dotted and dashed curves) 
      in the vicinity of the zero energy resonance at $B_0=155.041$ G. 
      While the dotted curve refers to the universal estimate, just depending
      on the scattering length $a$, the dashed curve also accounts for 
      the van der Waals interaction $-C_6/r^6$, at large inter-atomic 
      distances $r$, in terms of the mean scattering length $\bar{a}$ 
      \cite{Gribakin93,KGB03}. The dotted-dashed line indicates the linear 
      dependence of the energy $E_\mathrm{res}(B)$, associated with the 
      Feshbach resonance level $\phi_\mathrm{res}(r)$, on the magnetic field 
      strength. Here and throughout this paper, we have chosen the zero of the 
      energy, at each magnetic field strength $B$, as the energy of two 
      asymptotically free $^{85}$Rb atoms in the $(F=2,m_F=-2)$ state.}
    \label{fig:Eb85Rb}
  \end{center}
\end{figure}

The frequencies of the fringes have been identified as the vibrational 
frequencies of the highest excited diatomic vibrational 
bound state $\phi_\mathrm{b}(B_\mathrm{evolve})$ at the magnetic 
field strength $B_\mathrm{evolve}$ of the evolution period  
\cite{Donley02}, through their remarkable agreement with the result 
of coupled channels calculations (cf.~Fig.~\ref{fig:Eb85Rb}). 
%This agreement provided the main evidence for molecular production. 
The marked magnetic field dependence of the binding energy 
$E_\mathrm{b}(B_\mathrm{evolve})$ associated with the bound state 
$\phi_\mathrm{b}(B_\mathrm{evolve})$ is depicted in Fig.~\ref{fig:Eb85Rb}.
We note that 
the range of frequencies measured via bulk properties of the gas extends 
over three orders of magnitude from several kHz to the MHz regime. The 
associated oscillation periods are very much shorter than typical time scales 
set by the bulk motion of dilute Bose-Einstein condensates, which are 
typically on the order of the trap period of about 1/10 of a second in 
Refs.~\cite{Donley02,Claussen03}. 

For the regime of the high MHz oscillation 
frequencies the spatial extent of the molecular wave function is typically on 
the order of several nanometres \cite{TKTGPSJKB03}, 
while the length scale of the atomic cloud is set by the trap length of about 
3000 nm. In view of these enormous differences in the orders of magnitude the 
visibility of the Ramsey fringes is a remarkable phenomenon in its own right.
In the opposite regime of low kHz oscillation frequencies, i.e.~very close to 
the zero energy resonance at $B_0=155.041$ G, Claussen {\em et al.} 
\cite{Claussen03} have observed pronounced upward shifts of the fringe 
frequencies with respect to the vibrational frequencies of the molecules.
The frequency shifts were accompanied by a strong damping of the fringe 
pattern. This damping suggests an enhancement of the 
influence of many-body phenomena related to the coherent nature of the dilute 
Bose-Einstein condensate on the fringe pattern, as the magnetic field strength 
approaches the zero energy resonance. 
 
\subsection{Ramsey interferometry with atoms and molecules in a 
  Bose-Einstein condensate}
\label{subsec:Ramseyinterferometry}
%Before we provide the exact magnitudes of the oscillation frequency and  
%visibility of the fringe pattern in the remnant condensate fraction, in both 
%frequency regimes (cf.~Subsection 
%\ref{Frequencyandvisibilityoftheinterferencefringes} and
%Section \ref{sec:ExactoscillationfrequenciesoftheRamseyfringes}), 
%we shall identify the different elements of a usual Ramsey interferometer 
%from the pulse sequence in Fig.~\ref{fig:pulse}. 
We shall now identify the different elements of a Ramsey 
interferometer from the pulse sequence in Fig.~\ref{fig:pulse}. Our analysis
will show that in the regime of comparatively high molecular vibrational 
frequencies the atom-molecule oscillation experiment 
\cite{Donley02,Claussen03} 
can be directly interpreted in terms of Ramsey interferometry, while in 
the opposite low frequency regime a loss of coherence is expected to 
wipe out the fringes. 

\subsubsection{Interpretation of the magnetic field pulses in terms of beam 
  splitters}
We shall first show that the 
pulses $\#1$ and $\#2$ of Fig.~\ref{fig:pulse} provide the analogue of 
beam splitters through rapid tuning of the inter-atomic forces in the gas 
between the weakly and the strongly interacting regime. This is achieved by 
the variation of the scattering length $a(B)$, which characterises the static 
binary interactions in dilute Bose-Einstein condensates and depends on the 
magnetic field strength $B$ in accordance with the formula: 
\begin{align}
  a(B)=a_\mathrm{bg}
  \left(
  1-\frac{\Delta B}{B-B_0}
  \right).
  \label{aofB}
\end{align}
Here $a_\mathrm{bg}=-443\, a_\mathrm{Bohr}$ 
($a_\mathrm{Bohr}=0.052918$ nm is the Bohr radius) and 
$\Delta B=10.71\, \mathrm{G}$ \cite{Claussen03} are the background 
scattering length and the resonance width, respectively. At the minimum of the 
magnetic field pulses the stationary magnetic field strength of 
$B_\mathrm{min}=155.5$ G thus implies a large positive scattering length 
of about $10000\, a_\mathrm{Bohr}$, which is close to the mean inter-atomic 
distance of about $12000\, a_\mathrm{Bohr}$ in the dilute gas. 
The singularity of the scattering length in Eq.~(\ref{aofB}) is
a decisive feature of all zero energy resonances in ultra-cold collision 
physics and indicates the emergence of a highly excited diatomic vibrational 
bound state $\phi_\mathrm{b}(B)$ at the magnetic field strength 
$B_0$ of the zero energy resonance. This molecular bound state exists just on 
the side of positive scattering lengths (cf.~Fig.~\ref{fig:Eb85Rb}), and its 
mean internuclear distance is largely determined by $\langle r\rangle=a(B)/2$ 
\cite{Grisenti00,TKTGPSJKB03}. 

The favourable overlap between the diatomic energy states and the 
many-body state of the gas provides the necessary 
conditions for producing the desired strongly correlated pairs of atoms in the 
molecular bound state $\phi_\mathrm{b}(B_\mathrm{evolve})$, during the upward 
ramp at the end of the first magnetic field pulse. The association of bound
molecules at the final magnetic field strength $B_\mathrm{evolve}$ of 
Pulse $\# 1$ is accompanied by the production of correlated pairs of atoms 
in the continuum of positive energies above the two-body dissociation
threshold. The occupation of these continuum states provides a first burst 
component that diffuses during the evolution period.
The first magnetic field pulse thus splits the virtually uncorrelated 
initial Bose-Einstein condensate into three components; the remnant 
Bose-Einstein condensate, bound molecules and burst atoms.

\subsubsection{The evolution period}
In the ideal limit of coherent Ramsey interferometry, the gas is 
weakly interacting during the evolution period, 
i.e.~$B_\mathrm{evolve}$ is sufficiently far from the zero energy resonance 
that the scattering length $a(B)$ is much smaller than the mean distance 
between the atoms in the dilute gas. The remnant condensate, burst and 
bound molecular components are then virtually orthogonal and evolve 
coherently. 
%This also implies that bound molecules can at all be identified 
%as a separate entity of the gas \cite{TKTGPSJKB03}. 
The amplitude of the molecular component in the state 
$\phi_\mathrm{b}(B_\mathrm{evolve})$ thus acquires a phase shift of 
$\Delta \varphi=|E_\mathrm{b}(B_\mathrm{evolve})|t_\mathrm{evolve}/\hbar$ 
(cf.~Fig.~\ref{fig:Eb85Rb}) relative to the slowly varying amplitudes of the 
burst and condensate components. 

\subsubsection{The second magnetic field pulse}
The second magnetic field pulse gets all three components to overlap  
and thereby probes the phase differences between their amplitudes. The
final densities of the three components are therefore sensitive 
to the phases of their amplitudes at the end of the evolution period, which 
explains the emergence of a fringe pattern as a function of 
$t_\mathrm{evolve}$ as well as its angular oscillation frequency of 
$\omega_\mathrm{e}=|E_\mathrm{b}(B_\mathrm{evolve})|/\hbar$. 

We note that in order to observe coherent
Ramsey fringes, the different components of the system need to be orthogonal
during the evolution period. In Ramsey interferometers with light or
atomic beams the beam splitters provide the spatially separated, 
i.e.~orthogonal, components. In the atom-molecule coherence experiment 
orthogonality is provided by the ideally large difference in the orders of 
magnitude between the mean inter-atomic distance in the atomic burst and 
condensate components on one hand and the comparatively small bond length 
of the bound molecules during the evolution period on 
the other hand. In both cases the lack of overlap prevents any significant 
exchange between the components during the evolution period.

\subsection{Frequency and visibility of the interference fringes}
\label{Frequencyandvisibilityoftheinterferencefringes}
In this subsection we determine the visibility and oscillation frequency 
of the Ramsey fringes in the remnant Bose-Einstein condensate and provide
quantitative corroboration of the more physical arguments provided in 
Subsection \ref{subsec:Ramseyinterferometry}. We shall provide the main 
physical ideas of the derivations in terms of the dynamics of a single pair 
of atoms and present the results of a genuine many-body treatment to clearly 
separate two-body phenomena from those specifically related to the coherent 
nature of the partially Bose condensed gas. The detailed derivations on the 
basis of the many-body approach are given in Appendix 
\ref{app:Manybodydynamics}. We restrict our analysis in this section to the 
ideal limit of Ramsey interferometry, in which the gas is weakly interacting 
during the evolution period.

\subsubsection{Atom-molecule interference of a single pair of atoms}
Motivated by the approaches of Refs.~\cite{Mies00,Borca03}, we first consider 
a pair of $^{85}$Rb atoms in a spherically symmetric trap, exposed to a 
spatially homogeneous magnetic field, whose strength is varied in time, 
according to the sequence of pulses in Fig.~\ref{fig:pulse}. The degrees
of freedom of the centre of mass and relative motions of the atom pair 
are then exactly decoupled, and the dynamics of the relative coordinate 
$\mathbf{r}$ is determined by a Hamiltonian of the form: 
\begin{align}
  H_\mathrm{2B}=-\frac{\hbar^2}{m}\nabla^2+V_\mathrm{trap}(r)
  +V(B,r).
\end{align}
Here $V_\mathrm{trap}(r)$ is the potential of the atom trap, $m$ is the atomic
mass, and $V(B,r)$ is
the magnetic field dependent binary interaction potential. In general, 
$H_\mathrm{2B}$ depends on all hyperfine states of the atoms, which are 
strongly coupled by the Feshbach resonance state $\phi_\mathrm{res}(r)$ 
(cf., e.g., Refs.~\cite{Mies00,GKGTJ03} and Fig.~\ref{fig:Eb85Rb}). For
simplicity, we shall suppress, throughout this paper, all indices related to 
the multichannel nature of the binary interactions. All our results and 
derivations are, however, quite general and can be applied to zero energy 
resonances with any number of relevant scattering channels.

The analogue of the condensate mode is realised by the lowest energetic 
state of two atoms above the dissociation threshold of the pair interaction 
potential $V(B,r)$, whose wave function $\phi_0(r)$ thus obeys the 
stationary Schr\"odinger equation $H_\mathrm{2B}\phi_0(r)=E_0\phi_0(r)$, with 
the two-body Hamiltonian evaluated at the initial magnetic field strength of 
the pulse sequence. The amplitude for the probability to detect the atoms in 
the state $\phi_0$ at the final time $t_\mathrm{final}$ of the pulse sequence 
thus provides the analogue of the amplitude for the remnant condensate 
fraction, and is determined by:
\begin{align}
  T_\mathrm{2B}(\phi_0\leftarrow\phi_0)=
  \langle\phi_0|U_\mathrm{2B}(t_\mathrm{final},t_0)|\phi_0\rangle.
  \label{transamp2Bgeneral}
\end{align}
Here $U_\mathrm{2B}(t_\mathrm{final},t_0)$ is the time evolution operator, 
which satisfies the Schr\"odinger equation
\begin{align}
  i\hbar\frac{\partial}{\partial t}U_\mathrm{2B}(t,t_0)
  =H_\mathrm{2B}(t)U_\mathrm{2B}(t,t_0).
  \label{SEU2B}
\end{align}

Separating the pulse sequence in Fig.~\ref{fig:pulse} into its different 
components, $U_\mathrm{2B}(t_\mathrm{final},t_0)$ can be factorised into the 
evolution operators $U_{\# 1}(t_1,t_0)$, $U_{\# 2}(t_\mathrm{final},t_2)$ and 
$U_\mathrm{evolve}(t_\mathrm{evolve})$, associated with the first and second 
magnetic field pulse and with the evolution period of the sequence, 
respectively. Here $t_1$ is the time immediately after the first magnetic 
field pulse, $t_2$ is the time at the beginning of the second magnetic field 
pulse, and $t_2-t_1=t_\mathrm{evolve}$ is the duration of the evolution 
period. The factorisation of $U_\mathrm{2B}(t_\mathrm{final},t_0)$ yields:
\begin{align}
  %\nonumber
  U_\mathrm{2B}(t_\mathrm{final},t_0)
  %&=\mathcal{T}
  %\left[
  %  \exp
  %  \left(-\frac{i}{\hbar}
  %  \int_{t_0}^{t_\mathrm{final}}dt\, H_\mathrm{2B}(t)
  %  \right)
  %  \right]\\
  &=
  U_{\# 2}(t_\mathrm{final},t_2)U_\mathrm{evolve}(t_\mathrm{evolve})
  U_{\# 1}(t_1,t_0).
  \label{factorisationU2B}
\end{align}
During the evolution period the magnetic field strength 
is constant and $U_\mathrm{evolve}(t_\mathrm{evolve})$ depends only 
on the duration $t_\mathrm{evolve}$. It can therefore be written in terms of 
the stationary energy states of $H_\mathrm{2B}$ at the magnetic field strength 
$B_\mathrm{evolve}$ (cf.~Fig.~\ref{fig:pulse}). This leads to the spectral 
decomposition:
\begin{align}
  U_\mathrm{evolve}(t_\mathrm{evolve})=\sum_{v=-1}^{\infty}
  |\phi_v^\mathrm{evolve}\rangle 
  e^{-iE_v^\mathrm{evolve}t_\mathrm{evolve}/\hbar}
  \langle\phi_v^\mathrm{evolve}|.
  \label{Uevolvespectral}
\end{align}
Here we have labelled the two-body energy states $\phi_v^\mathrm{evolve}(r)$ 
by their vibrational quantum numbers, starting from the highest excited 
vibrational molecular bound state 
$\phi_{-1}^\mathrm{evolve}(r)=\phi_\mathrm{b}^\mathrm{evolve}(r)$, whose 
energy $E_\mathrm{b}(B_\mathrm{evolve})=E_\mathrm{b}^\mathrm{evolve}$ is 
depicted in Fig.~\ref{fig:Eb85Rb}, to the quasi continuum of excited states 
$\phi_{v}^\mathrm{evolve}(r)$ ($v=0,1,2,\ldots$) with energies above the 
dissociation threshold of $V(B_\mathrm{evolve},r)$. As the initial state, 
as well as the trap potential, are spherically symmetric, we have left out all 
energy states corresponding to partial waves with an orbital angular momentum 
different from $l=0$. We have also neglected all vibrational molecular bound 
states with energies below $E_\mathrm{b}(B_\mathrm{evolve})$ because these 
states are energetically inaccessible.

Inserting Eq.~(\ref{Uevolvespectral}) into Eq.~(\ref{factorisationU2B}), the 
transition amplitude in Eq.~(\ref{transamp2Bgeneral}) can be represented in
terms of a diffusion contribution $D_\mathrm{2B}(t_\mathrm{evolve})$, which 
is slowly varying in $t_\mathrm{evolve}$ and stems from the quasi continuum 
states, along with a comparatively rapidly varying oscillatory amplitude 
$A_\mathrm{2B}(t_\mathrm{evolve})$ associated with the spectral component of 
the molecular bound state:
\begin{align}
  T_\mathrm{2B}(\phi_0\leftarrow\phi_0)=D_\mathrm{2B}(t_\mathrm{evolve})
  +A_\mathrm{2B}(t_\mathrm{evolve}).
\end{align}
Both amplitudes can be expressed in terms of the amplitudes 
\begin{align}
  \label{transamp1}
  T_\mathrm{2B}^{\# 1}(\phi_v^\mathrm{evolve}\leftarrow\phi_0)
  =&\langle\phi_v^\mathrm{evolve}|U_{\# 1}(t_1,t_0)|\phi_0\rangle,
  %=\sqrt{P_{0,v}^{\# 1}}e^{i\varphi_{0,v}^{\# 1}},
  \\
  T_\mathrm{2B}^{\# 2}(\phi_0\leftarrow\phi_v^\mathrm{evolve})
  =&\langle\phi_0|U_{\# 2}(t_\mathrm{final},t_2)|\phi_v^\mathrm{evolve}\rangle
  %=\sqrt{P_{v,0}^{\# 2}}e^{i\varphi_{v,0}^{\# 2}},
  \label{transamp2}
\end{align}
associated with the probabilities for transitions between $\phi_0(r)$ and the 
vibrational states $\phi_v^\mathrm{evolve}(r)$ at the magnetic field strength 
$B_\mathrm{evolve}$ during the first and second magnetic field pulse. 
This gives the diffusion part to be
\begin{align}
  \nonumber
  D_\mathrm{2B}(t_\mathrm{evolve})=&
  \sum_{v=0}^{\infty}
  T_\mathrm{2B}^{\# 2}(\phi_0\leftarrow\phi_v^\mathrm{evolve})
  e^{-iE_v^\mathrm{evolve}t_\mathrm{evolve}/\hbar}\\
  &\times
  T_\mathrm{2B}^{\# 1}(\phi_v^\mathrm{evolve}\leftarrow\phi_0),
\end{align}
while the oscillatory amplitude is given by
\begin{align}
  \nonumber
  A_\mathrm{2B}(t_\mathrm{evolve})=&
  T_\mathrm{2B}^{\# 2}(\phi_0\leftarrow\phi_\mathrm{b}^\mathrm{evolve})
  e^{-iE_\mathrm{b}^\mathrm{evolve}t_\mathrm{evolve}/\hbar}
  \\
  &\times
  T_\mathrm{2B}^{\# 1}(\phi_\mathrm{b}^\mathrm{evolve}\leftarrow\phi_0).
\end{align}

The probability $P_{0,0}(t_\mathrm{evolve})=
\left|T_\mathrm{2B}(\phi_0\leftarrow\phi_0)\right|^2$
for the two atoms to be detected in the state $\phi_0(r)$ at the final time 
of the pulse sequence provides the two-body analogue of the remnant 
condensate fraction as a function of $t_\mathrm{evolve}$. Neglecting the
weak dependence of the diffusion term $D_\mathrm{2B}(t_\mathrm{evolve})$
on $t_\mathrm{evolve}$, $P_{0,0}(t_\mathrm{evolve})$ takes the well known 
form of an interference signal:
\begin{align}
  \nonumber
  P_{0,0}(t_\mathrm{evolve})
  =&
  %\left|
  %T_\mathrm{2B}(\phi_0\leftarrow\phi_0)
  %\right|^2
  %\nonumber
  %=&
  \left|D_\mathrm{2B}\right|^2
  +\left|A_\mathrm{2B}\right|^2\\
  &+2\left|D_\mathrm{2B}\right|
  \sqrt{P_\mathrm{0,b}^{\# 1}P_\mathrm{b,0}^{\# 2}}
  \sin(\omega_\mathrm{e}t_\mathrm{evolve}+
  \Delta\varphi).
  \label{signal2B}
\end{align}
Here 
\begin{align}
  \label{transprob1}
  P_\mathrm{0,b}^{\# 1}=&\left|T_\mathrm{2B}^{\# 1}
  (\phi_\mathrm{b}^\mathrm{evolve}\leftarrow\phi_0)
  \right|^2, \\
  P_\mathrm{b,0}^{\# 2}=&\left|T_\mathrm{2B}^{\# 2}(\phi_0\leftarrow\phi_
  \mathrm{b}^\mathrm{evolve})\right|^2
  \label{transprob2}
\end{align}
are the probabilities for molecular formation and dissociation during
the first and second magnetic field pulse,
%associated with the amplitudes in Eqs.~(\ref{transamp1}) and 
%(\ref{transamp2}),
respectively, and 
\begin{align}
  \omega_\mathrm{e}=\left|E_\mathrm{b}^\mathrm{evolve}\right|/\hbar
  \label{frequency2B}
\end{align}
is the angular frequency of the fringes. The total phase shift $\Delta\varphi$ 
involves the phases of the complex amplitudes in Eqs.~(\ref{transamp1}) and 
(\ref{transamp2}), and of $D_\mathrm{2B}$.

The signal in Eq.~(\ref{signal2B}) has maxima and minima at the evolution 
times $t_\mathrm{evolve}^\mathrm{max}$ and $t_\mathrm{evolve}^\mathrm{min}$,
which fulfil the relations
\begin{align}
  \label{tevolvemax}
  \sin\left(\omega_\mathrm{e} t_\mathrm{evolve}^\mathrm{max}+
  \Delta\varphi\right)=&1,\\
  \sin\left(\omega_\mathrm{e} t_\mathrm{evolve}^\mathrm{min}+
  \Delta\varphi\right)=&-1,
  \label{tevolvemin}
\end{align}
respectively. The visibility $V_\mathrm{Ramsey}$ of the Ramsey fringes 
produced by a single pair of atoms is thus given by:
\begin{align}
  \nonumber
  V_\mathrm{Ramsey}
  &=\frac{P_{0,0}(t_\mathrm{evolve}^\mathrm{max})-
    P_{0,0}(t_\mathrm{evolve}^\mathrm{min})}
  {P_{0,0}(t_\mathrm{evolve}^\mathrm{max})+
    P_{0,0}(t_\mathrm{evolve}^\mathrm{min})}\\
  &=
  2\sqrt{P_\mathrm{0,b}^{\# 1}P_\mathrm{b,0}^{\# 2}/P_{0,0}^\mathrm{avg}}.
  \label{VRamsey2B}
\end{align}
Here $P_{0,0}^\mathrm{avg}=|D_\mathrm{2B}|^2+|A_\mathrm{2B}|^2$ is the 
average signal obtained for evolution times 
$t_\mathrm{evolve}$, at which the oscillatory part on the right hand side of 
Eq.~(\ref{signal2B}) vanishes. 

These derivations based on two-body physics reveal that the emergence of the 
Ramsey fringes in Refs.~\cite{Donley02,Claussen03} does not crucially rely 
upon the presence of many atoms in a Bose-Einstein condensate. The same kind 
of interferometry could be realised, for instance, also with two atoms in a 
tight atom trap of an optical lattice site. These atoms could be bosons or 
non-identical fermions, which both can interact with each other via $s$ waves. 
The visibility of the fringe pattern is determined by the probability 
$P_\mathrm{0,b}^{\# 1}$ of molecular production in the first magnetic field 
pulse and by the probability $P_\mathrm{b,0}^{\# 2}$ of their reconversion
into atom pairs in the lowest energetic quasi continuum state $\phi_0(r)$,
during the second magnetic field pulse. The angular frequency of the Ramsey 
fringes is solely determined by the binding energy 
$E_\mathrm{b}^\mathrm{evolve}$.

\subsubsection{Continuum limit}
To extend the two-body considerations to an effective description of a 
homogeneous gas of atoms, we shall first describe the limiting process as
we move from the discrete trap states to a continuum. We thus assume that the 
two atoms are confined to a large volume $\mathcal{V}$, whose length scales 
greatly exceed the range of the inter-atomic forces as well as the spatial 
extent of the bound molecular state $\phi_\mathrm{b}^\mathrm{evolve}$. The 
products $\mathcal{V}P_\mathrm{0,b}^{\# 1}$ and 
$\mathcal{V}P_\mathrm{b,0}^{\# 2}$ are then virtually independent of 
$\mathcal{V}$, and the coupling between the centre of mass and the relative 
coordinates of the atom pair is negligible. In the continuum limit, the 
amplitudes 
$T_\mathrm{2B}^{\# 1}(\phi_\mathrm{b}^\mathrm{evolve}\leftarrow\phi_0)$ 
and $T_\mathrm{2B}^{\# 2}(\phi_0\leftarrow\phi_\mathrm{b}^\mathrm{evolve})$
%on the right hand sides of Eqs.~(\ref{transprob1}) and (\ref{transprob2}) 
contain the two-body time evolution operator and
the bound molecular state $\phi_\mathrm{b}^\mathrm{evolve}$ in free space,
as well as the limit of the lowest energetic quasi continuum 
state $|\phi_0\rangle$ to a zero momentum plane wave $|0\rangle$ of the 
relative motion of the atoms. This implies the formal replacement:
\begin{align}
  |\phi_0\rangle\to|0\rangle\sqrt{\frac{(2\pi\hbar)^3}{\mathcal{V}}}
   \, .
\end{align}
This choice of the initial state properly reflects the coherence properties of a Bose-Einstein
condensate \cite{Stoof89}.

In contrast to molecular production or dissociation, the transfer between 
quasi continuum states involves a spatial range on the order of magnitude of
the complete sample volume $\mathcal{V}$. This prevents us from performing the 
continuum limit directly in the elastic amplitude 
$T_\mathrm{2B}(\phi_0\leftarrow\phi_0)$ of 
Eq.~(\ref{transamp2Bgeneral}). We consider instead the continuum limit of the 
function
\begin{align}
  \nonumber
  h(t,t_0)&=\mathcal{V}i\hbar\frac{\partial}{\partial t}
  \langle\phi_0|U_\mathrm{2B}(t,t_0)|\phi_0\rangle\\
  &\to
  (2\pi\hbar)^3\langle0|V(B(t))U_\mathrm{2B}(t,t_0)|0\rangle\theta(t-t_0),
  \label{couplingfunction}
\end{align}
which is related to the elastic transition amplitude through the following:
\begin{align}
  T_\mathrm{2B}(\phi_0\leftarrow\phi_0)=
  1-\frac{i}{\hbar\mathcal{V}}\int_{t_0}^{t_\mathrm{final}}
  dt\, h(t,t_0).
  \label{elasticampcontinuum}
\end{align}
Here $\theta(t-t_0)$ denotes the step function, which yields unity for 
$t>t_0$ and zero elsewhere. The first term on the right hand side of
Eq.~(\ref{elasticampcontinuum}) can to some extent be interpreted as 
the part of the amplitude associated with the events in which both atoms do 
not scatter, while the second term, proportional to the inverse of the volume, 
accounts for all possible collision events during the pulse sequence.

\subsubsection{Probabilistic extension of the description of a single pair 
  of atoms to the description of a homogeneous gas}
A dilute gas usually occupies a volume $\mathcal{V}$ much larger than 
the volume needed for a single pair of atoms to collide and then separate 
to a distance outside the range of their binary interaction. The volume of the 
gas can thus be divided into regions that are small enough to contain at most
two atoms but still sufficiently large to exceed the range of the inter-atomic 
forces by a large amount. This coarse graining procedure provides the 
physical justification for rescaling two-body transition amplitudes in order 
to describe a homogeneous gas of atoms with a density $n$. The correct scaling 
factor between the effective volume of two colliding atoms and the entire 
volume $\mathcal{V}$ of the gas can be determined from classical probability 
theory. When applied to the molecular conversion efficiency of the first 
magnetic field pulse, for instance, classical probability theory predicts the 
number $2N_\mathrm{b}^{\# 1}$ of atoms, associated to molecules, relative to 
the total number $N=n\mathcal{V}$ of atoms in the gas to be:   
\begin{align}
  \nonumber
  2\frac{N_\mathrm{b}^{\# 1}}{N}
  &=
  \frac{2}{N}\frac{N(N-1)}{2}
  \frac{(2\pi\hbar)^3}{\mathcal{V}}
  \left|
  \langle\phi_\mathrm{b}|U_{\# 1}(t_1,t_0)|0\rangle
  \right|^2\\
  &=
  n\mathcal{V}P_\mathrm{0,b}^{\# 1}.
\end{align}
Here $2/N$ is the number of atomic constituents of a single diatomic molecule 
relative to 
the total number of atoms, $N(N-1)/2\approx N^2/2$ is the number of pairs of 
atoms in the gas, and the remaining factor $(2\pi\hbar)^3
\left|\langle\phi_\mathrm{b}|U_{\# 1}(t_1,t_0)|0\rangle\right|^2/\mathcal{V}$
is the probability for the association of a single pair of atoms in the 
continuum limit. The probabilistic extension of the description of a single 
pair of atoms to the description of a homogeneous gas thus consists in 
identifying the effective volume of a colliding pair of atoms by the inverse 
of the atomic density. This identification is quite intuitive, as $1/n$ is 
exactly the volume, in which, on average, a single atom can be found in a 
homogeneous gas. We thus replace the two-body amplitude on the right hand 
sides of Eq.~(\ref{elasticampcontinuum}) by 
\begin{align}
  T_\mathrm{2B}^\mathrm{gas}(\phi_0\leftarrow\phi_0)=
  1-n\frac{i}{\hbar}\int_{t_0}^{t_\mathrm{final}}
  dt\, h(t,t_0).
  \label{T00MB}
\end{align}

The probabilistic approach implies, in particular, that we should multiply 
the transition probabilities for molecular production or dissociation in 
Eqs.~(\ref{transprob1}) and (\ref{transprob2}) by the number of atoms 
$N=n\mathcal{V}$. A similar procedure was previously applied by 
Mies {\em et al.} (cf.~Refs.~\cite{Mies00,GKGTJ03}), leading to a remarkable
agreement between theory and the Feshbach resonance crossing experiments of
Ref.~\cite{Stenger99}. From its derivation given in this paper, however, it 
becomes apparent that the probabilistic extension of the two-body 
physics to the description of a gas is not restricted to Bose-Einstein 
condensates but could be applied also to the interacting pairs of atoms in 
dilute two component Fermi gases, boson-fermion mixtures or non-condensed Bose 
gases. A comparison between Eq.~(\ref{T00MB}) and the dynamic equation 
(\ref{NMNLSE}) for the atomic mean field of the genuine many-body theory of 
Refs.~\cite{Koehler02,KGB03} implies that the applicability of the 
probabilistic extension of the two-body physics is restricted to time scales 
$t_\mathrm{final}-t_0\ll m/[2\hbar a(B_\mathrm{evolve}) 
n_\mathrm{c}^\mathrm{evolve}]$, where $n_\mathrm{c}^\mathrm{evolve}$ is the 
mean condensate density during the evolution period, 
and $2\pi\hbar/\mu_\mathrm{evolve}=m/[2\hbar a(B_\mathrm{evolve}) 
n_\mathrm{c}^\mathrm{evolve}]$ is the oscillation period associated with the 
chemical potential.   

\begin{figure}[htb]
  \begin{center}
    \includegraphics[width=\columnwidth,clip]{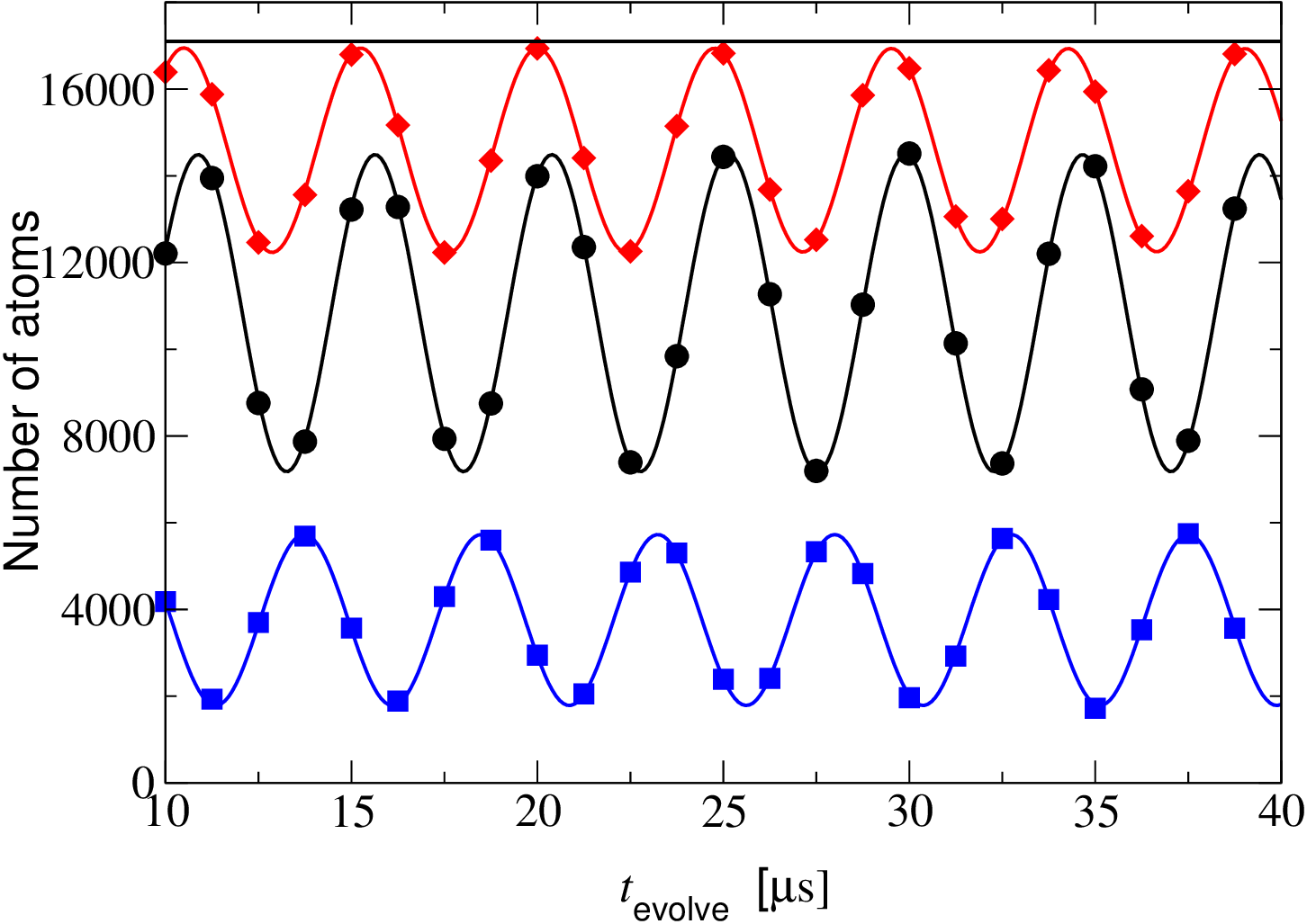}
    \caption{Number of remnant condensate and burst atoms in dependence on 
      $t_\mathrm{evolve}$, as determined from the probabilistic extension 
      of the two-body dynamics, evaluated at the mean experimental density 
      \cite{Donley02} of $3.16\times 10^{12}\, \mathrm{cm}^{-3}$. All the 
      other physical parameters and symbols are the same as those in 
      Fig.~\protect\ref{fig:85RbAMcoherence}.} 
    \label{fig:85RbAMcoherence2-body}
  \end{center}
\end{figure}

Figure \ref{fig:85RbAMcoherence2-body} shows the interference fringes 
predicted by the probabilistic extension of the two-body dynamics for the 
experimental situation described in Fig.~6 of Ref.~\cite{Donley02}. The 
comparison between Fig.~\ref{fig:85RbAMcoherence} of this paper and Fig.~6 of 
Ref.~\cite{Donley02} on one hand and Fig.~\ref{fig:85RbAMcoherence2-body} on 
the other hand reveals that the two-body approach overestimates 
molecular production efficiencies and the fringe visibilities in a 
Bose-Einstein condensate. The overall qualitative features, as well as the 
order of magnitude of all components, however, correctly reflect the 
predictions of full many-body simulations as well as the experimental 
observations. The predictive power of such a simple approach based on 
classical probability theory is even more remarkable in view of the fact that 
the widely used two level mean field approach of
Refs.~\cite{Drummond98,Timmermans99,Yurovsky00,Salgueiro01,Adhikari01,Cusack02,Abdullaev03} 
fails completely to describe the fringe pattern. In fact, this version of a
many-body mean field theory not only rules out the production of 
burst atoms \cite{GKGTJ03} but it also predicts, when applied to the physical 
situation of Fig.~6 in Ref.~\cite{Donley02}, that the number of 
condensate atoms should remain virtually unchanged, throughout the entire 
pulse sequence. This failure of the two level mean field approach is due to 
its incomplete description of the binary collision physics \cite{GKGTJ03}.

Equation (\ref{VRamsey2B}) implies that the visibility of the 
interference fringes in the remnant Bose-Einstein condensate is given by
\begin{align}
  V_\mathrm{Ramsey}=2n\mathcal{V}
  \sqrt{P_\mathrm{0,b}^{\# 1}P_\mathrm{b,0}^{\# 2}/P_{0,0}^\mathrm{avg}},
  \label{VRamseyextended}
\end{align}
where $P_{0,0}^\mathrm{avg}$ can be obtained from the transition amplitude in 
Eq.~(\ref{T00MB}) and thus converges to unity in the limit of zero density. 
The product $\mathcal{V}\sqrt{P_\mathrm{0,b}^{\# 1}P_\mathrm{b,0}^{\# 2}}$ is 
independent of the density $n$ and of the volume $\mathcal{V}$. Consequently,
the visibility $V_\mathrm{Ramsey}$ varies linearly with the density in the 
limit $n\to 0$. The probabilistic approach recovers the angular 
frequency of the Ramsey fringes of the associated two-body problem in 
Eq.~(\ref{frequency2B}).

\subsubsection{Molecular production efficiency}
The visibility of the Ramsey fringes in Eq.~(\ref{VRamseyextended}) strongly 
depends on the molecular production efficiency $P_\mathrm{0,b}^{\# 1}$,
during the first magnetic field pulse of the sequence in Fig.~\ref{fig:pulse}.
In the course of the molecular association, pairs of atoms separated by the 
average inter-atomic distance of the gas are transferred into a highly excited
but comparatively tightly bound state, with a mean internuclear distance on 
the order of $\langle r\rangle=a(B_\mathrm{evolve})/2$. These length scales 
typically differ by several orders of magnitude, and special techniques are 
required to efficiently overcome the problem of the large spatial separation 
of atoms in a dilute gas. 

The technique of magnetic field pulses, applied in 
Refs.~\cite{Donley02,Claussen03}, uses two steps to achieve molecular 
association. First, the magnetic field strength is swept close to the zero 
energy resonance, so that the scattering length $a(B_\mathrm{min})$ becomes 
comparable in magnitude to the mean inter-atomic distance of the gas. From the 
intermediate length scale
set by $a(B_\mathrm{min})$ the atom pairs are then transferred to their
mean internuclear distance of $\langle r\rangle=a(B_\mathrm{evolve})/2$
in the final upward ramp of the first magnetic field pulse. This physical 
picture suggests that an efficient molecular association requires a 
favourable balance
between the mean inter-atomic distance of the gas and the scattering lengths 
$a(B_\mathrm{min})$ and $a(B_\mathrm{evolve})$. A similar balance of length 
scales is also required to achieve favourable Franck-Condon overlaps between
initial and final molecular states in the general context of molecular 
spectroscopy \cite{Bransden03}. Our analytic studies of 
Appendix \ref{app:Molecularproduction}, indeed, indicate that the mechanism of 
molecular production with magnetic field pulses exhibits analogies to the 
experimental technique of two colour photo-association
(see Ref.~\cite{Julienne98} for a description of the technique), in which 
free pairs of atoms are transferred into a comparatively tightly bound 
molecule via Raman transitions through a highly excited intermediate state.
While the photo-association approach can populate a single deeply bound target 
level, by appropriately tuning the laser frequencies, the molecular production 
via magnetic field pulses involves a range of energies that overlap with 
the continuum part of the two-body spectrum. This reduces the efficiency of 
the association of molecules in the highest excited diatomic vibrational 
bound state and leads to the observed production of burst atoms
\cite{Donley02}.

\begin{figure}[htb]
  \begin{center}
    \includegraphics[width=\columnwidth,clip]{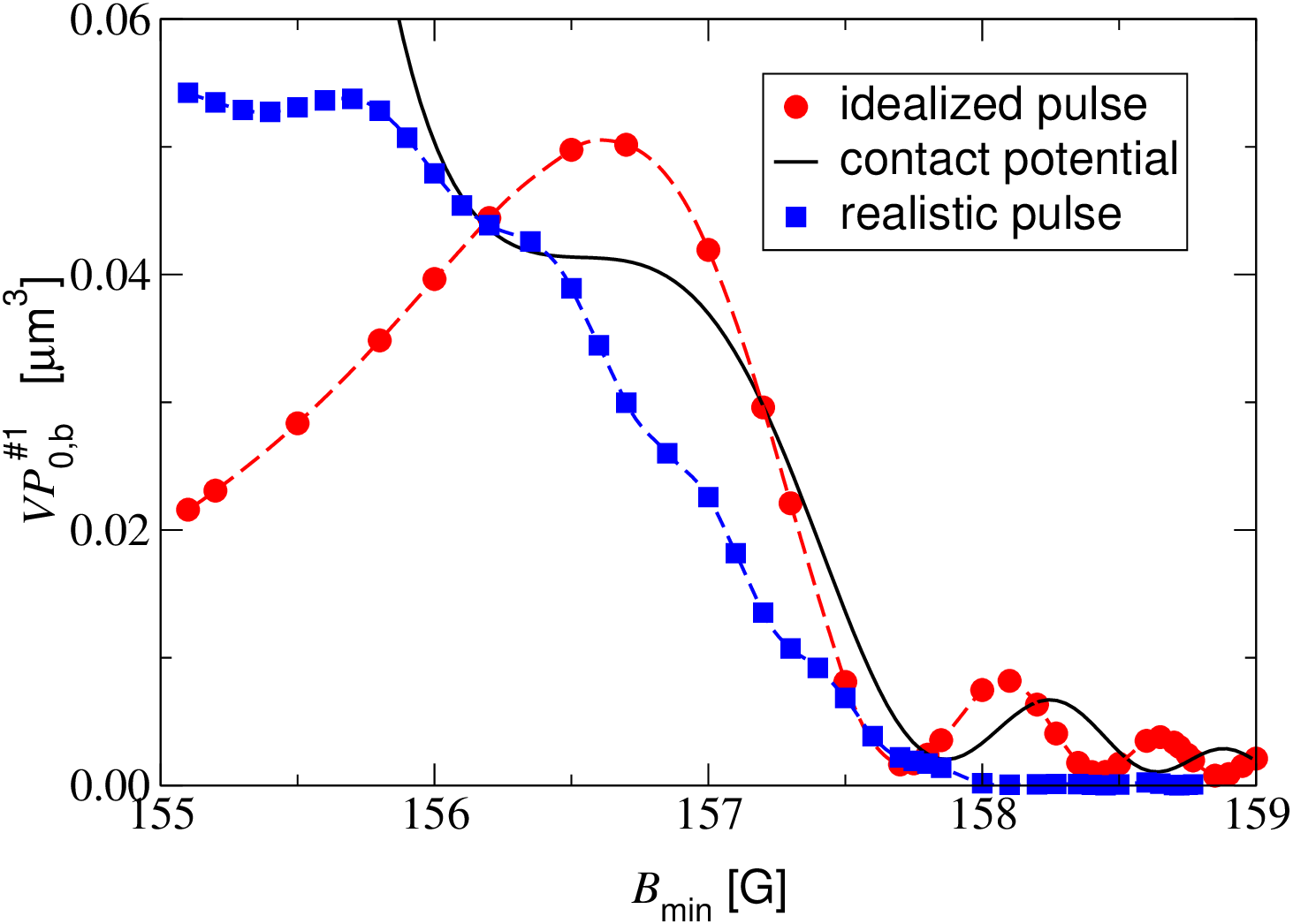}
    \caption{The molecular production efficiency $P_\mathrm{0,b}^{\# 1}$
      as a function of the minimum magnetic field strength 
      $B_\mathrm{min}$, during the first pulse of the sequence of 
      Fig.~\protect\ref{fig:pulse}, for a fixed evolution field strength of 
      $B_\mathrm{evolve}=160$ G. The figure compares molecular production 
      efficiencies of the realistic pulse of Fig.~\protect\ref{fig:pulse} 
      (squares) with predictions associated with an idealised pulse sequence
      (cf.~Appendix \ref{app:Manybodydynamics}),
      which just contains sudden changes of the magnetic field strength. These 
      predictions involve calculations using a realistic low energy binary
      interaction potential \cite{KGB03,KGG03,GKGTJ03} (circles and squares) 
      and the contact potential approximation of 
      Appendix \ref{app:Molecularproduction} (solid curve). The duration of
      the idealised first magnetic field pulse of the sequence was chosen to 
      be 36.2 $\mu$s.}
    \label{fig:efficiency1hump}
  \end{center}
\end{figure}

These considerations are confirmed by the results on the molecular
production efficiency in the first pulse of the magnetic field
sequence presented in Fig.~\ref{fig:efficiency1hump}. For the case of
an idealised pulse sequence, consisting of sudden jumps of the magnetic
field strength, we observe an optimal value of $B_\mathrm{min}$ of
around 156.7 G, where a favourable balance between the three inter-particle
length scales involved in this process is reached. The analogous calculation
performed for the realistic pulse sequence determines the optimal value of
$B_\mathrm{min}$ to be smaller than 155.7 G, in agreement with the
pulse sequence actually employed in the experiments
of Refs.~\cite{Donley02,Claussen03}. Although both the realistic and
idealised pulse sequence calculations predict the same order of magnitude for 
the molecular production efficiency, they significantly differ in their
precise dependences on the magnetic field $B_\mathrm{min}$. This
reveals the sensitivity of the molecular component produced, and, as a
consequence, of the observed fringe visibility, to the precise shape
of the magnetic field pulse used. Figure \ref{fig:efficiency1hump} also
shows that the analytic estimate of the molecular production
efficiency in the contact potential approximation for the idealised
pulse sequence (see Appendix \ref{app:Molecularproduction} for
details) recovers both the magnitude and the essential features of the
predictions based on a realistic low energy binary interaction
potential. For the values of $B_\mathrm{min}$ smaller than 156 G
we are outside the domain of the analytical approximation, as
explained in Appendix \ref{app:Molecularproduction}. For the idealised
pulse sequence we observe effects of quantum interference in the
inelastic transition amplitude resulting in the oscillations of the
transition probability for $B_\mathrm{min}$ greater than 157 G.
These interference phenomena in inelastic transition amplitudes are sometimes 
referred to as St\"uckelberg oscillations \cite{Child74}
(see, e.g., Ref.~\cite{Esry99} for 
similar interferences in three-body recombination rate constants).

\subsubsection{Many-body corrections to the frequency and 
  visibility of the Ramsey fringes in the remnant Bose-Einstein 
  condensate}
The main differences between the effective two-body and the genuine many-body 
descriptions of dilute Bose-Einstein condensates reside in the treatment
of the coherent nature of the gas and of the non-linear dynamics of the atomic 
mean field. In Appendix \ref{app:Manybodydynamics} we provide a detailed 
derivation of the many-body corrections to the frequency and visibility 
of the Ramsey fringes in the remnant Bose-Einstein condensate, under the 
assumption that the gas is weakly interacting during the evolution period of 
the pulse sequence. In this section we give a physical explanation of their 
origin. For simplicity, we shall consider all quantities in the homogeneous 
gas limit.

The coherent dynamics of the atomic mean field $\Psi(t)$,
during the evolution period, is largely described by 
the stationary limit
$\Psi(t)\propto 
(n_\mathrm{c}^\mathrm{evolve})^{1/2}\exp(-i\mu_\mathrm{evolve}t/\hbar)$,
up to a global phase shift, where $n_\mathrm{c}^\mathrm{evolve}$ is the mean 
condensate density and
\begin{align} 
  \mu_\mathrm{evolve}=4\pi\hbar^2a(B_\mathrm{evolve})
  n_\mathrm{c}^\mathrm{evolve}/m 
  \label{muGPE}
\end{align}
is the chemical potential. The chemical 
potential can thus be largely interpreted as the amount of energy gained by 
each atom due to the presence of the surrounding atoms. We thus expect that
the finite energy of $2\mu_\mathrm{evolve}$ in a binary collision leads to 
an upward shift of the two-body angular fringe frequency in 
Eq.~(\ref{frequency2B}), which yields:  
\begin{align}
  \omega_\mathrm{e}=\left(\left|E_\mathrm{b}^\mathrm{evolve}\right|+
  2\mu_\mathrm{evolve}\right)/\hbar.
  \label{omegaeManyBody}
\end{align}
This is precisely the estimate that we obtain from the systematic treatment  
given in Appendix \ref{app:Manybodydynamics}. 

We note, however, that the Gross-Pitaevskii approximation of the chemical 
potential in Eq.~(\ref{muGPE}) becomes singular when the magnetic field 
strength approaches the zero energy resonance. This singularity is an artifact 
of the assumption of weak interactions, 
i.e.~$n_\mathrm{c}^\mathrm{evolve}a^3(B_\mathrm{evolve})\ll 1$,  
during the evolution period. The Gross-Pitaevskii 
approximation of Eq.~(\ref{muGPE}) may be systematically improved using the 
self consistent treatment of Ref.~\cite{KGG03}, which determines the chemical 
potential in terms of the exact stationary solution of the genuine many-body 
dynamic equation (\ref{NMNLSE}) for the atomic mean field
\cite{footnote:stationary}. This stationary approach leads to a chemical
potential, which properly accounts for the energy dependence of the binary
transition amplitudes \cite{KGG03} and leads to the physically expected 
smooth behaviour of $\mu_\mathrm{evolve}$ when the magnetic field strength
is tuned in the close vicinity of the zero energy resonance. 

Our studies of 
Section \ref{sec:ExactoscillationfrequenciesoftheRamseyfringes}
also indicate that even in the absence of deeply inelastic loss processes,
like, e.g., spin relaxation, the experimental \cite{Donley02,Claussen03} 
evolution times $t_\mathrm{evolve}$ are too short for the gas to equilibrate. 
In particular, $t_\mathrm{evolve}$ is typically much smaller than the time
scale $\hbar/(2\mu_\mathrm{evolve})$ set by the many-body frequency shift 
in Eq.~(\ref{omegaeManyBody}). As a consequence, the experiments 
\cite{Donley02,Claussen03} could not fully resolve the frequency shift.

The derivations in Appendix \ref{app:Manybodydynamics} reveal that the 
many-body corrections to the visibility in Eq.~(\ref{VRamseyextended}) are 
mainly due to the non-linear dynamics of the atomic mean field. The 
final result  
\begin{align}
  V_\mathrm{Ramsey}=\frac{1-\exp
    \left(
    -4n\mathcal{V}
    \left[
      P_{0,\mathrm{b}}^{\#1}P_{\mathrm{b},0}^{\#2}
      \right]^{1/2}
    \right)}
  {1+\exp
    \left(
    -4n\mathcal{V}
    \left[
      P_{0,\mathrm{b}}^{\#1}P_{\mathrm{b},0}^{\#2}
      \right]^{1/2}
    \right)}
  \label{VisibilityManyBody}
\end{align}
largely recovers the dependences of Eq.~(\ref{VRamseyextended}) on the 
physical parameters $n$, $\mathcal{V}P_{0,\mathrm{b}}^{\#1}$
and $\mathcal{V}P_{\mathrm{b},0}^{\#2}$, so that both formulae
asymptotically coincide in the limit $n\to 0$. The non-linear nature of the
mean field dynamics, however, leads to a reduction of the fringe visibility
in comparison with the probabilistic extension of the linear two-body 
time evolution, when the density is increased. A similar behaviour can be 
observed in the comparison between genuine many-body predictions and the 
extended two-body Landau-Zener approach of Ref.~\cite{Mies00} for the 
description of the molecular production in Feshbach resonance crossing 
experiments with linear ramps of the magnetic field strength \cite{GKGTJ03}.  

\section{Decoherence of the Ramsey fringes}
\label{sec:ExactoscillationfrequenciesoftheRamseyfringes}
In this section we consider in detail the regime of magnetic field strengths
in the close vicinity of the zero energy resonance, in which decoherence  
phenomena diminish the Ramsey fringes in the remnant 
Bose-Einstein condensate. This regime of low molecular vibrational frequencies 
is strongly influenced by 
%a contribution of binary collisions and 
the non-linear dynamics of the atomic mean field. Our numerical simulations 
are therefore based on the many-body approach of Refs.~\cite{Koehler02,KGB03} 
that includes both a non-perturbative treatment of the binary physics and the 
non-linearities to all orders in the atomic mean field. 
  
\subsection{Comparison between experimental Ramsey fringes in the remnant
  Bose-Einstein condensate and many-body simulations}
In the first order microscopic quantum dynamics approach of 
Refs.~\cite{Koehler02,KGB03} the evolution of all components of the gas is 
determined by a single non-linear Schr\"odinger equation for the atomic
mean field, i.e. 
\begin{align}
  \nonumber
  i\hbar\frac{\partial}{\partial t}\Psi(\mathbf{x},t)=& 
  \left[-\frac{\hbar^2}{2m}\nabla^2+V_\mathrm{trap}(\mathbf{x})\right]
  \Psi(\mathbf{x},t)\\
  &-\Psi^*(\mathbf{x},t)
  \int_{t_0}^\infty d\tau \
  \Psi^2(\mathbf{x},\tau)\frac{\partial}{\partial \tau}h(t,\tau).
  \label{NMNLSE}
\end{align}
The inter-atomic collisions are included fully in this approach in
the time dependent coupling function of Eq.~(\ref{couplingfunction}). 
The dynamics of the atomic mean field determines not only the condensate 
density $n_\mathrm{c}(\mathbf{x},t)=|\Psi(\mathbf{x},t)|^2$ but also the
molecular and burst fractions of the gas \cite{KGB03}. The simulations of 
the final number of atoms of all components of the gas in 
Fig.~\ref{fig:85RbAMcoherence} correspond to the experimental situation of 
Fig.~6 in Ref.~\cite{Donley02}, and were obtained from a sequence of full 
solutions of Eq.~(\ref{NMNLSE}) with variable evolution times.

In the course of our studies, we have solved Eq.~(\ref{NMNLSE}) 
for a variety of experimental \cite{Claussen03} evolution times 
$t_\mathrm{evolve}$ and magnetic field strengths $B_\mathrm{evolve}$.
Our simulations fully account for realistic sequences of magnetic field 
pulses \cite{Hodby}, shown schematically in Fig.~\ref{fig:pulse},
and for the inhomogeneous nature of the Bose-Einstein condensate in a 
cylindrical atom trap with the experimental radial and axial frequencies of 
$\nu_\mathrm{radial}=17.5$ Hz and $\nu_\mathrm{axial}=6.8$ Hz, respectively. 
The initial state of our simulations is the ground state of the stationary 
Gross-Pitaevskii equation, corresponding to a $^{85}$Rb Bose-Einstein 
condensate with $16000$ atoms \cite{Claussen03} at the initial magnetic field 
strength of $B=162.2$ G of the pulse sequence in Fig.~\ref{fig:pulse}. 

\begin{figure}[htb]
  \begin{center}
    \includegraphics[width=\columnwidth,clip]{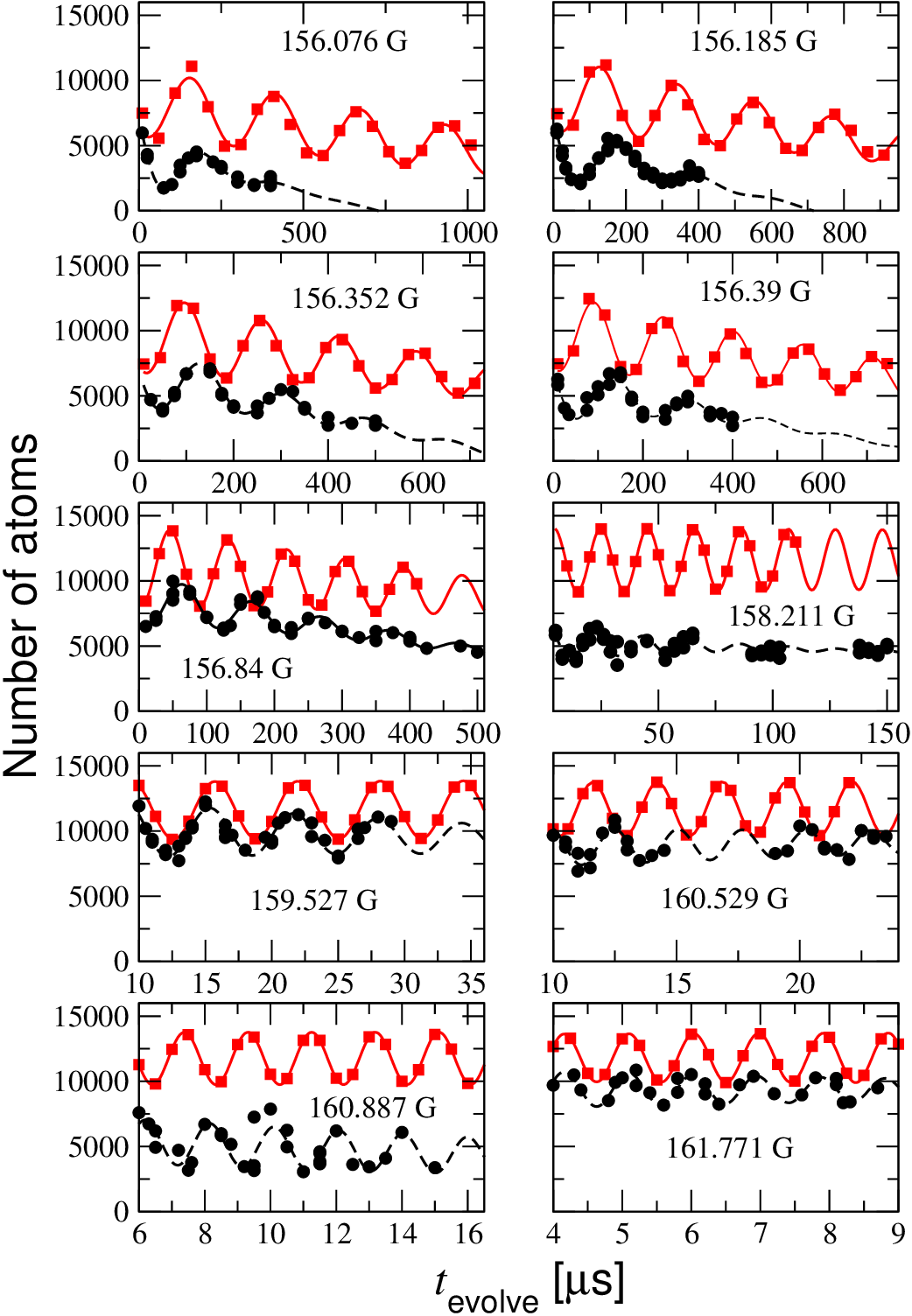}
    \caption{Comparison of the experimental \cite{Hodby,Claussen03} number of 
      atoms in the remnant Bose -Einstein condensate as a function of 
      $t_\mathrm{evolve}$ (circles) with the predictions of Eq.~(\ref{NMNLSE}) 
      (squares) at the end of each magnetic field pulse sequence. The dashed 
      and solid curves indicate the fits of Eq.~(\ref{fit}) \cite{Claussen03}
      to the experimental and theoretical data, respectively. In the course of 
      the experiments \cite{Claussen03}, the magnetic field strength 
      $B_\mathrm{evolve}$ during the evolution period of the pulse sequence
      was varied between 156.076 and 161.771 G. The fringe patterns cover
      a wide range of evolution times, from about $10\, \mu$s at 161.771 G 
      to about 1 ms at 156.076 G.}
    \label{fig:stamp}
  \end{center}
\end{figure}

Figure \ref{fig:stamp} shows comparisons between the experimentally observed 
dependence on $t_\mathrm{evolve}$ of the remnant condensate and 
theoretical predictions of Eq.~(\ref{NMNLSE}), at the end of each magnetic 
field pulse sequence. Both the experimental data and the theoretical 
predictions exhibit damped oscillations, whose decay becomes particularly 
marked for sequences with longer evolution times on the order of 1 ms, at 
magnetic field strengths of $B_\mathrm{evolve}\lesssim 157$ G, i.e.~in the 
close vicinity of the zero energy resonance at $B_0=155.041$ G 
\cite{Claussen03}. The theoretical predictions have the same systematic trends 
as the experimental fringe patterns with frequencies between 4 kHz 
at 156.076 G and about 1 MHz at 161.771 G. All of the theoretical 
curves overestimate the measured numbers of remnant condensate atoms. These 
deviations may, at least in principle, be due to fast inelastic 
loss phenomena such as spin relaxation in $^{85}$Rb collisions, whose rates 
are enhanced in the vicinity of the zero energy resonance \cite{Roberts00}. 
We shall show in Subsection \ref{subsec:twobodydecay}, however, that spin 
relaxation is unlikely to play a dominant role in these deviations. Our 
results of Subsection \ref{Frequencyandvisibilityoftheinterferencefringes}
rather indicate that the remnant condensate fractions in the interferometric 
experiments are very sensitive to the precise shape of the magnetic field 
pulse sequence. We therefore believe that a mismatch between the real 
experimental pulse sequence and the sequence used in our simulations 
is a more likely reason for the deviations, in particular, for magnetic field 
strengths $B_\mathrm{evolve}\gtrsim 158$ G, away from the position of the
zero energy resonance.

\subsection{Oscillation frequencies of the Ramsey fringes in the remnant 
  Bose-Einstein condensate} 
In order to extract the fringe frequencies from our simulations, we have 
closely followed the procedure suggested in Ref.~\cite{Claussen03} and fitted 
the theoretically predicted dependence of the number of remnant condensate 
atoms on $t_\mathrm{evolve}$ to the formula:
\begin{align}
  \nonumber
  N_\mathrm{c}^\mathrm{evolve}=&
  N_\mathrm{avg} 
  -\alpha t_\mathrm{evolve}\\ 
  &
  +A e^{-\beta t_\mathrm{evolve}}
  \sin{(\omega_\mathrm{e} t_\mathrm{evolve}+
    \Delta\varphi)}.
  \label{fit}
\end{align}
The angular frequency $\omega_\mathrm{e}$ of the fringes has been interpreted 
in Ref.~\cite{Claussen03} in terms of a natural oscillation frequency $\nu_0$, 
which is related to the angular frequency $\omega_\mathrm{e}$ through 
$\omega_\mathrm{e}=2\pi\sqrt{\nu_0^2-[\beta/(2\pi)]^2}$. The amplitude 
$A$ of the interference fringes as well as the average number $N_\mathrm{avg}$
of atoms determine the fringe visibility. The remaining fit parameters are a 
loss rate of atoms $\alpha$, a damping rate $\beta$ and a phase shift 
$\Delta\varphi$ of the Ramsey fringes. We note that the fit parameters of
Eq.~(\ref{fit}) can be largely identified from the interference signal of a 
single pair of atoms in Eq.~(\ref{signal2B}), provided that the weak 
dependence of the diffusion contribution $D_\mathrm{2B}(t_\mathrm{evolve})$ on 
the evolution time $t_\mathrm{evolve}$ is taken into account. The fit at 
$B_\mathrm{evolve}=156.076$ G in Fig.~\ref{fig:stamp} indicates, however, 
that the functional form of Eq.~(\ref{fit}) is not entirely compatible with
the non-linear dynamics of the atomic mean field in Eq.~(\ref{NMNLSE}). 

\begin{figure}[htb]
  \begin{center}
    \includegraphics[width=\columnwidth,clip]{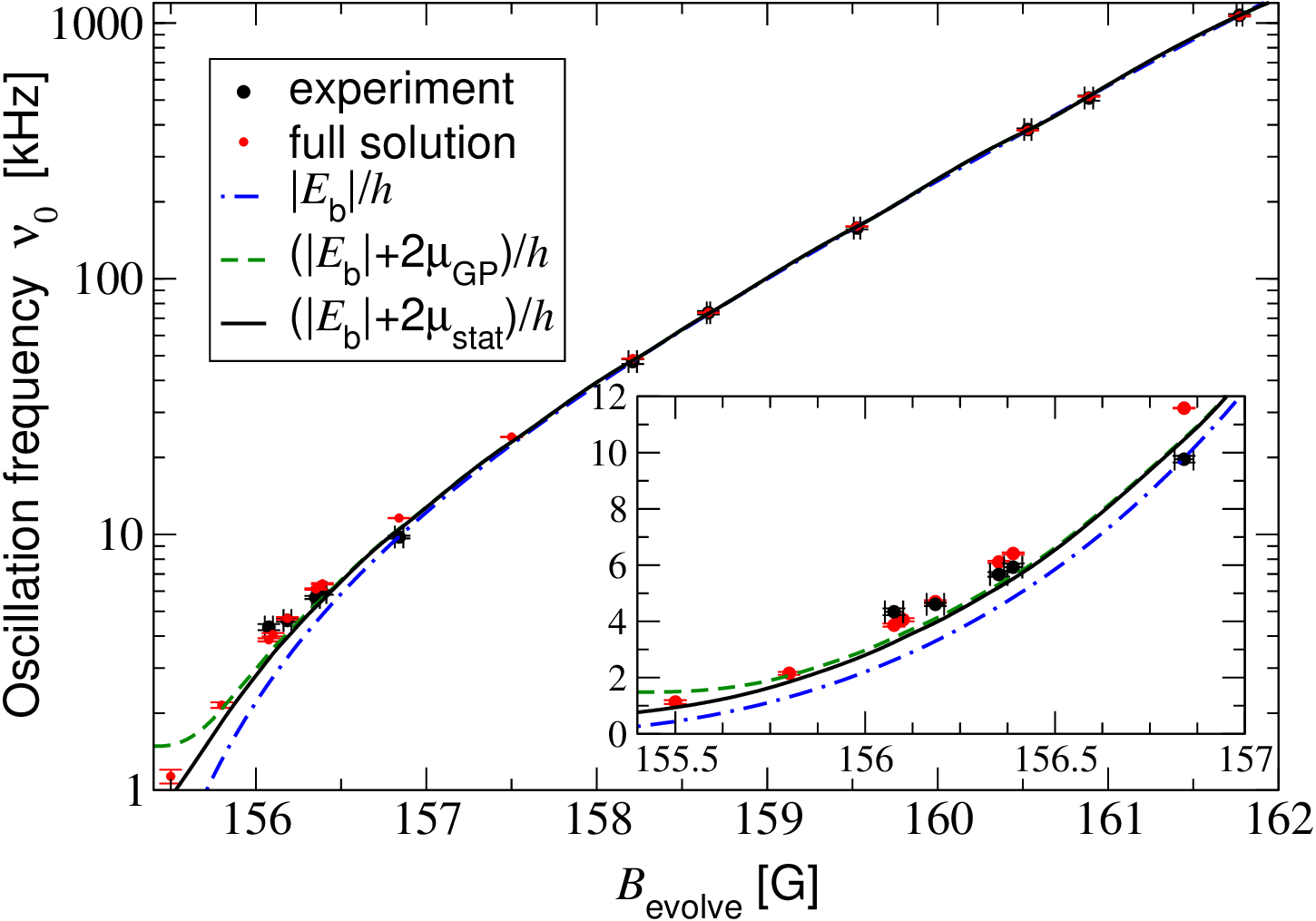}
    \caption{Natural frequency $\nu_0$ of the Ramsey fringes as a
      function of the magnetic field strength $B_\mathrm{evolve}$ as 
      obtained through fits of the theoretical predictions (squares)
      and of the experimental \cite{Hodby,Claussen03} number of remnant 
      condensate atoms (circles) to Eq.~(\ref{fit}). The solid and dashed
      curves indicate theoretical predictions of the fringe frequency that 
      account for the first order many-body contribution 
      $2\mu_\mathrm{evolve}/h$ 
      (cf.~Subsection \ref{Frequencyandvisibilityoftheinterferencefringes}) 
      of the chemical potential in the self 
      consistent stationary approach of Ref.~\protect\cite{KGG03}
      (solid curve) and in the Gross-Pitaevskii approximation of 
      Eq.~(\ref{muGPE}) (dashed curve), respectively. The 
      dotted dashed curve shows the binding frequency of the highest excited 
      vibrational molecular bound state of $^{85}$Rb$_2$, in dependence on 
      the magnetic field strength. The inset shows an enlargement of the 
      near resonance region of magnetic field strengths $B_\mathrm{evolve}$, 
      in which the deviations between fringe frequency and molecular 
      vibrational frequency become apparent. Small error bars in 
      the theoretical data result from the uncertainties associated with 
      the fitting procedure.}
    \label{fig:freq}
  \end{center}
\end{figure}

Figure \ref{fig:freq} shows the natural frequency $\nu_0$ in dependence on 
$B_\mathrm{evolve}$ as obtained from the fits to the experimental and
theoretical data in Fig.~\ref{fig:stamp}, respectively. As pointed out in 
Refs.~\cite{Donley02,Kokkelmans02,KGB03}, for magnetic field strengths
$B_\mathrm{evolve}>158$ G, sufficiently far away from the zero energy 
resonance at $B_0=155.041$ G, the fringe frequencies closely
match the vibrational frequencies associated with the weakly bound molecular 
state $\phi_\mathrm{b}^\mathrm{evolve}$. A comparison with 
Fig.~\ref{fig:Eb85Rb} shows that the binding energies in this range of 
magnetic field strengths are sensitive to the long range van der Waals 
dispersion coefficient $C_6$ \cite{KGB03} and even to the magnetic moment 
$dE_\mathrm{res}/dB$ of the Feshbach resonance level 
(cf.,e.g., Refs.~\cite{Mies00,GKGTJ03}), which determines the 
linear slope of $E_\mathrm{b}(B)$ at magnetic field strengths 
$B_\mathrm{evolve}>162$ G.

Closer to the zero energy resonance
both the experimental data \cite{Claussen03} as well as the theoretical 
predictions show a visible upward shift of the fringe frequencies from the 
molecular vibrational frequencies. From the analytic results of our 
theoretical studies in Section \ref{sec:PhysicaloriginoftheRamseyfringes} 
and from our numerical simulations
we expect this shift to persist for all values of $B_\mathrm{evolve}$ but 
to decrease in its relative magnitude with increasing magnetic field 
strengths. The four experimental points closest to the zero energy resonance, 
to which a shift has been attributed in the experimental reference 
\cite{Claussen03}, agree in their magnitudes with our theoretical predictions 
(see inset of Fig.~\ref{fig:freq}). For magnetic field strengths 
$B_\mathrm{evolve}\geq 156.84$ G the shift was neglected in 
Ref.~\cite{Claussen03} and the associated fringe frequencies were used to 
adjust the theoretical curve of the binding frequencies in 
Figs.~\ref{fig:Eb85Rb} and \ref{fig:freq}. This, in turn, also determines the 
parameters $B_0$, $\Delta B$ and $a_\mathrm{bg}$ of Eq.~(\ref{aofB}).
As we have used these parameters as an input to the coupling function 
$h(t,\tau)$ of Eq.~(\ref{NMNLSE}), and the frequency shift persists for
all magnetic field strengths $B_\mathrm{evolve}$ in our approach, we expect
our simulations to deviate slightly from the experimental findings simply due
to the slight uncertainty in the Feshbach resonance parameters.

\subsection{Two-body decay of the Bose-Einstein condensate}
\label{subsec:twobodydecay}
Despite the fact that the fit parameter $\alpha$ of Eq.~(\ref{fit}) is 
determined with a much larger uncertainty than the frequency $\nu_0$, we
may interpret it in terms of two-body decay processes 
during the evolution period. Two-body loss phenomena in a 
Bose-Einstein condensate are quite generally described by the rate equation
\begin{align}
  \dot{N}_\mathrm{c}(t)=-\frac{K_2}{2} \langle n_\mathrm{c}(t)\rangle
  N_\mathrm{c}(t).
  \label{twobodyrateequation}
\end{align}
Here $K_2$ is the two-body loss rate constant, 
$\langle n_\mathrm{c}(t)\rangle$ is the average condensate density, and
$N_\mathrm{c}(t)$ is the number of condensate atoms. We note that 
Eq.~(\ref{twobodyrateequation}) is not restricted to the description of 
spin relaxation events but refers to all two-body loss 
phenomena which occur on time scales much shorter than those set by the
bulk motion of the gas. We shall restrict our considerations on the two-body 
decay to magnetic field strengths $B_\mathrm{evolve}< 157$ G
in the close vicinity of the zero energy resonance, where we find that 
the pulse sequences are sufficiently long to provide reliable values for 
$\alpha$ with a systematic trend as a function of $B_\mathrm{evolve}$. In 
this region the Ramsey fringes of Fig.~\ref{fig:stamp} also exhibit a strong 
decoherence, and diatomic molecules in the highest excited vibrational bound 
state can not be identified as a separate entity because their wave functions 
would be more extended than the mean inter-atomic distance of the gas 
\cite{TKTGPSJKB03}. In order to establish the relationship between $\alpha$ 
and $K_2$, we solve Eq.~(\ref{twobodyrateequation}) formally by dividing both 
sides by $N_\mathrm{c}(t)$ and then integrating the logarithmic derivative 
from the time $t_1$, immediately after the first magnetic field pulse of 
Fig.~\ref{fig:pulse}, to the time $t_2$, immediately after the evolution 
period. This yields: 
\begin{align}
  N_\mathrm{c}(t_2)=N_\mathrm{c}(t_1)
  \exp\left(-\frac{K_2}{2}\int_{t_1}^{t_2}dt\, 
  \langle n_\mathrm{c}(t)\rangle\right).
  \label{twobodyrateformalsol}
\end{align}
If we assume that during the evolution period $t_\mathrm{evolve}=t_2-t_1$
the loss of condensate atoms is small in comparison with $N_\mathrm{c}(t_1)$,
we can expand the right hand side of Eq.~(\ref{twobodyrateformalsol})
to first order in $t_\mathrm{evolve}$. This leads to the following expression:
\begin{align}
  N_\mathrm{c}(t_2)\approx N_\mathrm{c}(t_1)-
  \frac{K_2}{2}N_\mathrm{c}(t_1)\langle n_\mathrm{c}(t_1)\rangle 
  t_\mathrm{evolve}.
  \label{twobodyratelinear}
\end{align}
A comparison between Eqs.~(\ref{fit}) and (\ref{twobodyratelinear}) then 
allows us to identify the parameters $N_\mathrm{c}(t_1)=N_\mathrm{avg}$,
$N_\mathrm{c}(t_2)=N_\mathrm{c}^\mathrm{evolve}$ and 
\begin{align}
  K_2=\frac{2\alpha}{N_\mathrm{avg}\langle n_\mathrm{c}(t_1)\rangle},
  \label{K2fit}
\end{align}
provided
that the second magnetic field pulse does not significantly change the number 
of condensate atoms. Our numerical simulations show that this assumption
is justified for longer pulse sequences with evolution fields 
$B_\mathrm{evolve} < 157$ G (cf.~Fig.~\ref{fig:stamp}), while for 
the faster sequences it may be violated. 

\begin{figure}[htb]
  \begin{center}
    \includegraphics[width=\columnwidth,clip]{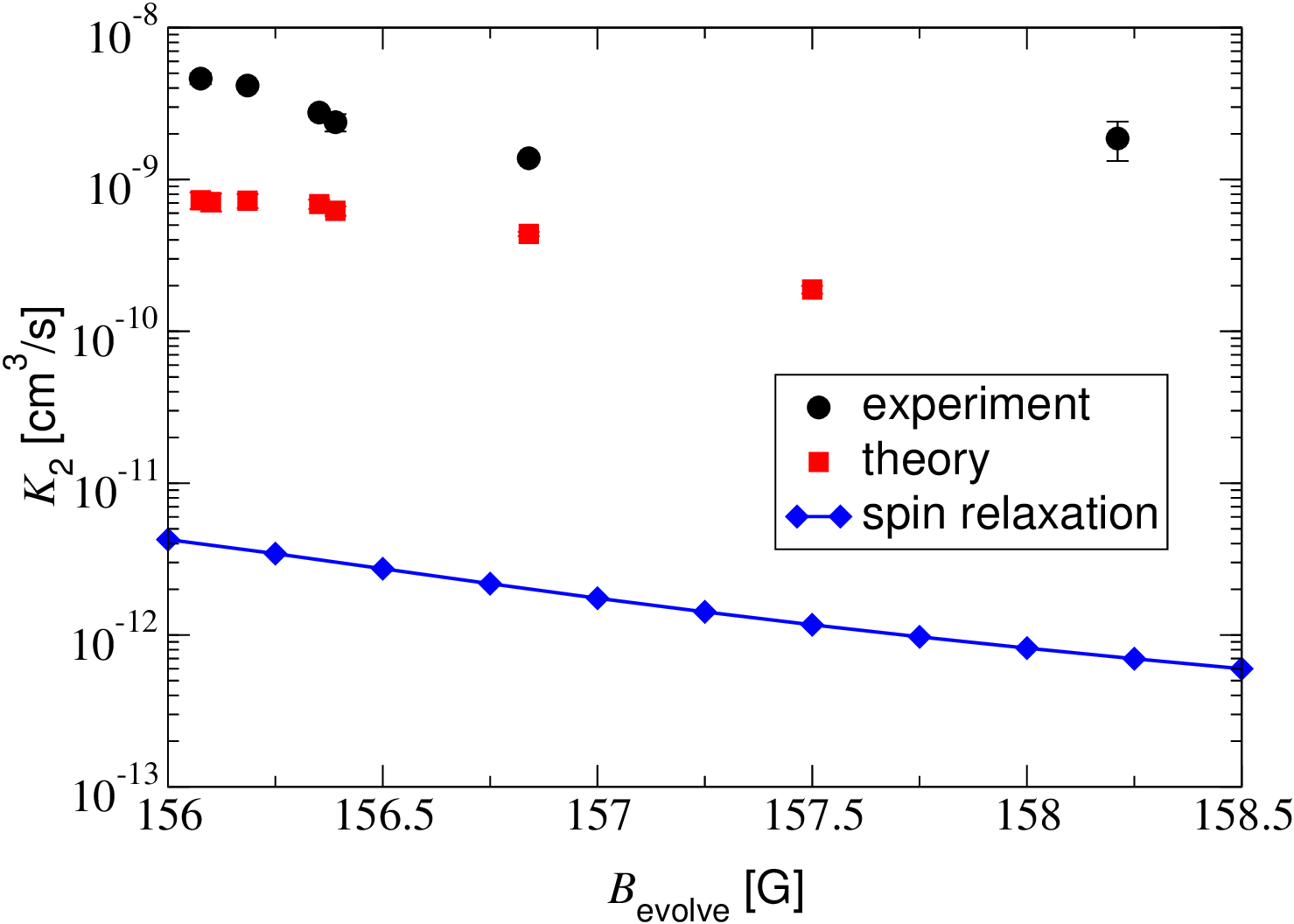}
    \caption{The two-body loss rate constant $K_2$ determined from 
    the fit parameter $\alpha$ of Eq.~(\ref{fit}) through Eq.~(\ref{K2fit}) 
    as a function of the 
    magnetic field strength $B_\mathrm{evolve}$ during the evolution period
    of the pulse sequence. The circles indicate the experimental values
    \cite{Claussen03}, while the squares correspond to the theoretical 
    predictions of Eq.~(\ref{NMNLSE}) depicted in Fig.~\ref{fig:stamp}.
    The error bars indicate the statistical errors of the fits. The solid 
    curve indicates coupled channels calculations \cite{Koehler04} of the spin 
    relaxation rate constant for a pair of $^{85}$Rb atoms in the limit of 
    zero collision energy. The rate constant $K_2$ is given on a logarithmic 
    scale.} 
    \label{fig:85Rbdecay2body}
  \end{center}
\end{figure}

Figure \ref{fig:85Rbdecay2body} provides an overview of the magnitudes of
the two-body loss rate constants $K_2$ obtained through Eq.~(\ref{K2fit}) 
from the fits to the experimental data \cite{Claussen03} (circles) and to the 
theoretical predictions (squares) of Fig.~\ref{fig:stamp}. The rate constants 
$K_2$ are shown as a function of the magnetic field strength 
$B_\mathrm{evolve}$ during the evolution period. 
For comparison, the solid curve indicates results of a coupled channels 
calculation \cite{Koehler04} of the spin relaxation rate constant. The 
comparatively large size of about $10^{-9}$ cm$^3$/s of both the experimental 
\cite{Claussen03} and theoretical rate constants 
strongly indicate that the loss of condensate atoms is mainly due to the 
energy transfer from the first magnetic field pulse. This drives the 
initially weakly interacting Bose-Einstein condensate into a strongly 
correlated non-equilibrium state. Deeply inelastic spin relaxation loss 
phenomena are not included in our simulations of Eq.~(\ref{NMNLSE}). The 
theory thus predicts that the atoms are transferred from the condensate into 
correlated pairs of atoms in excited states. The coupled channels calculations 
\cite{Koehler04}, indeed, confirm that the rate constants for spin relaxation 
(solid curve in Fig.~\ref{fig:85Rbdecay2body}) are less than 
$10^{-11}$ cm$^3$/s in the relevant range of magnetic field strengths, 
i.e.~from 156 to 158.5 G, and thus do not explain the size 
of the observed losses. 
%The large magnitudes of the loss rate constants in both the 
%experimental data \cite{Claussen03} and the theoretical predictions thus 
%strongly indicate that in the range of magnetic field strengths 
%$B_\mathrm{evolve} < 157$ G, close to the zero energy resonance, 
%the gas is far away from equilibrium and atoms are lost from the condensate 
%into the burst fraction \cite{Donley02} of atoms, while deeply inelastic spin 
%relaxation loss phenomena are unlikely to play a major role in the 
%experiments of Refs.~\cite{Donley02,Claussen03}. 
Our interpretation is corroborated by the observation of 
Ref.~\cite{Claussen03} that over the whole range of magnetic
field strengths $B_\mathrm{evolve}$ the loss rate $\alpha$ is weakly dependent 
on $B_\mathrm{evolve}$ and consistent with the number loss in a single 
magnetic field pulse \cite{Claussen02}.

\subsection{Visibility of the Ramsey fringes in the remnant Bose-Einstein 
  condensate}

\begin{figure}[htb]
  \begin{center}
    \includegraphics[width=\columnwidth,clip]{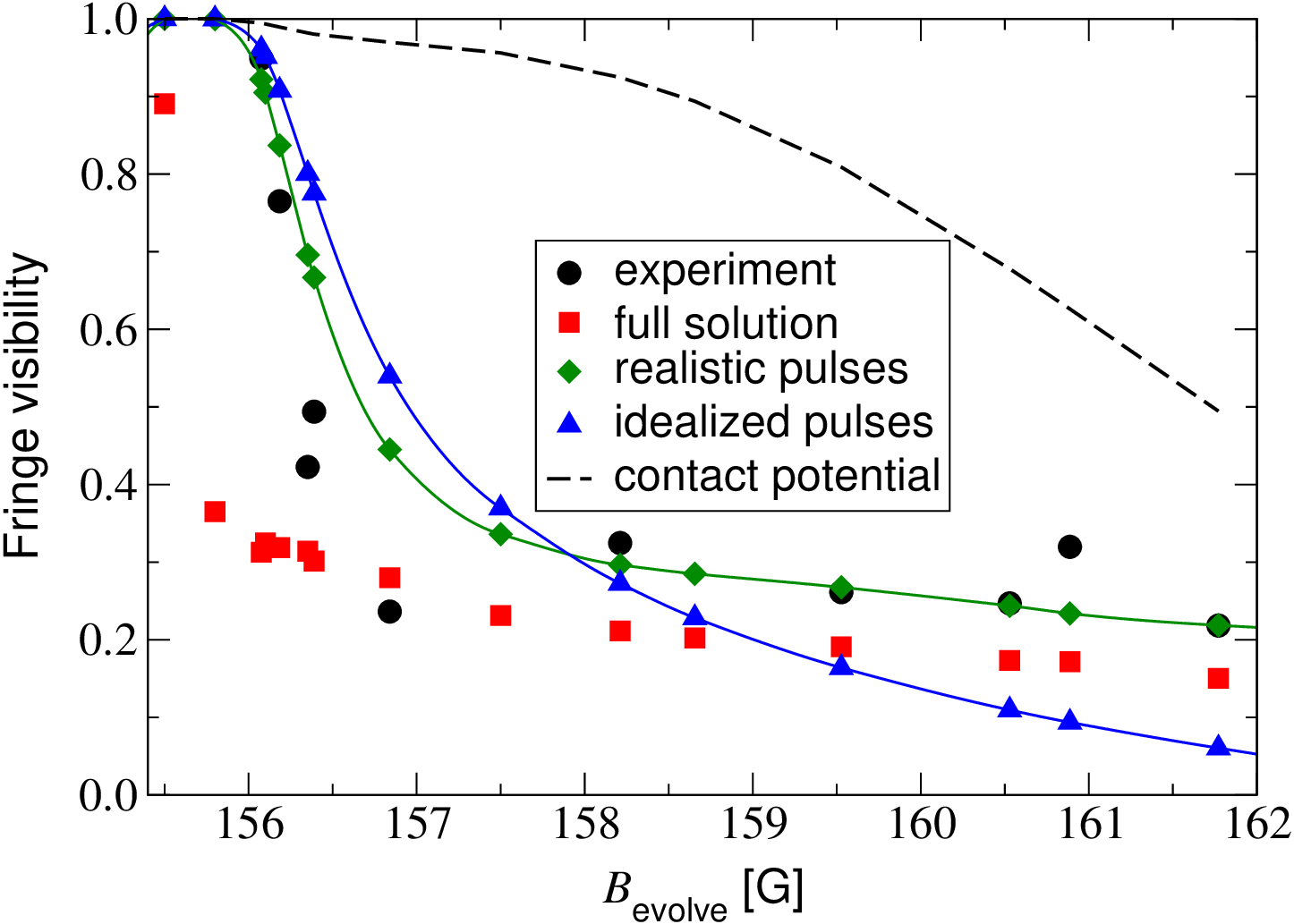}
    \caption{Visibility of the Ramsey fringes in the remnant Bose-Einstein 
      condensate determined from the experimental data (circles) and from the 
      solutions of Eq.~(\ref{NMNLSE}) (squares), as a function of the magnetic
      field strength $B_\mathrm{evolve}$ during the evolution period of the
      pulse sequence. We have also plotted the estimate of the fringe 
      visibility given by Eq.~(\ref{VisibilityManyBody}), evaluated for the 
      magnetic field pulses of the realistic sequence (diamonds) and of an 
      idealised pulse sequence with sudden changes of the magnetic field 
      strength (cf.~Appendix \ref{app:Manybodydynamics}) (triangles). For the 
      idealised pulse sequence the durations of the first pulse and the second 
      pulse were chosen to be 36.2 $\mu$s and 23.9 $\mu$s, respectively. 
      Equation (\ref{VisibilityManyBody}) has then been evaluated 
      for $n=3.07\times 10^{12}\, \mathrm{cm}^{-3}$, which corresponds to the
      initial mean density of the condensate in the experiments of 
      Ref.~\cite{Claussen03}. The dashed curve shows the predictions of the
      contact potential approximation to the probabilities 
      $P_{0,\mathrm{b}}^{\#1}$ and $P_{\mathrm{b},0}^{\#2}$ in 
      Eq.~(\ref{VisibilityManyBody}).}
    \label{fig:visibility}
  \end{center}
\end{figure}

Figure \ref{fig:visibility} shows the fringe visibility in the remnant 
Bose-Einstein condensate. This is determined as a function of the magnetic 
field strength $B_\mathrm{evolve}$ in the evolution period from the 
experimental data and from the full solutions of Eq.~(\ref{NMNLSE}). 
The visibilities were obtained from the fits of Eq.~(\ref{fit}) 
to the experimental and numerical data of Fig.~\ref{fig:stamp} 
using the relation
\begin{align}
  V_\mathrm{Ramsey}=A/N_\mathrm{avg}.
  \label{visibilityfit}
\end{align}
The theoretical results follow the rough size and monotonic trend of the 
experimental visibilities. The increase of 
$V_\mathrm{Ramsey}$ with decreasing $B_\mathrm{evolve}$ is due to the 
enhanced depletion of the average number $N_\mathrm{avg}$ of remnant 
condensate atoms at magnetic field strengths close to the zero energy 
resonance. This trend can also be observed in the visibilities estimated from
Eq.~(\ref{VisibilityManyBody}). These estimates were obtained from the exact
transition probabilities $P_{0,\mathrm{b}}^{\#1}$ and $P_{\mathrm{b},0}^{\#2}$
for both the pulses of the realistic sequence (as depicted in
Figure \ref{fig:pulse}) and their counterparts in an idealised pulse
sequence of sudden changes of the magnetic field strength
(cf.~Appendix \ref{app:Manybodydynamics}). Figure \ref{fig:visibility} also 
shows an estimate of $V_\mathrm{Ramsey}$ based on 
Eq.~(\ref{VisibilityManyBody}) and the contact potential approximation 
(cf.~Appendix \ref{app:Molecularproduction}). The contact potential 
approximation accounts for the inter-atomic interactions just in terms of a 
single parameter, the scattering length $a(B)$. The large difference between
the contact potential approximation and the experimental data as 
well as the exact theoretical data reveals the sensitivity of the 
interferometric experiments of Refs.~\cite{Donley02,Claussen03} to the 
nature of inter-atomic interactions beyond universal considerations, which are based 
solely on the scattering length. 
This sensitivity of the Ramsey interferometry with atoms and molecules
\cite{Donley02,Claussen03} to non-universal properties of the binary 
collision physics provides a crucial probe of the microscopic physics through 
the bulk properties of the gas. The information it provides is largely 
inaccessible to experimental studies involving linear ramps of the magnetic 
field strength across zero energy resonances \cite{KGG03,GKGTJ03}.    

\section{Conclusions}
Our theoretical studies in this paper reveal the remarkable range of physical 
regimes that are probed by the atom-molecule coherence experiments of 
Refs.~\cite{Donley02,Claussen03}. These regimes include the Ramsey 
interferometer limit, at evolution fields sufficiently far away from the zero 
energy resonance for the binary and bulk motions to be virtually decoupled.
In the other limit, i.e.~the close vicinity of the zero energy resonance, 
two- and many-body phenomena can not be disentangled during the evolution 
period. 

Using a probabilistic extension of the dynamics of a single pair of atoms to 
the description of a gas, we have clearly identified the different elements
of a Ramsey interferometer from the experimental \cite{Donley02,Claussen03}
sequences of magnetic field pulses. We have shown that much of 
the observations of Refs.~\cite{Donley02,Claussen03}, in particular,
the frequency and visibility of the Ramsey fringes, can be understood 
qualitatively and also largely quantitatively on the basis of 
two-body physics. Our studies reveal that the probabilistic extension of the 
pair dynamics is applicable provided that the experiments are operated in the 
interferometer mode, i.e.~at evolution fields far away from the zero 
energy resonance. These studies were motivated by the 
approaches of Refs.~\cite{Mies00,Borca03}. Our probabilistic approach, 
however, is not restricted to Bose-Einstein condensates and does not include
any adjustable parameters.

We have shown that in the interferometer mode, in contradiction to the 
interpretations of some previous theoretical studies 
\cite{Kokkelmans02,Mackie02,Duine03PRL}, the 
gas is virtually unaffected by non-equilibrium phenomena during the evolution 
period of the pulse sequence, and the mutually orthogonal molecular and free 
atomic components evolve independently and coherently. The Ramsey fringes 
thus reflect the phase difference of both components rather than dynamical 
exchange of particles between them. In fact, such an exchange between the 
molecular and free atomic components during the evolution period would imply 
the breakup and reformation of molecules. In the absence of any external 
driving force these would be energetically impossible.

In the Ramsey interferometer mode   
the measurements provide detailed insight into the binary collision dynamics
beyond universal considerations. These interferometric experiments are thus
sensitive to the long range van der Waals dispersion coefficient $C_6$ of the 
binary interactions and even to the magnetic moment of the Feshbach resonance 
level, rather than just to the scattering length. Our studies clearly show 
that this sensitivity is observed not only in the fringe frequencies but 
also in their visibilities. The existence of the fringe pattern and 
its dependence on the physical parameters of the system is remarkable in its 
own right with the experiments clearly showing how vibration of molecules on a
nanometre length scale can significantly influence the dynamics of the entire 
gas on length scales as large as several micrometres. This access to the 
microscopic physics can not easily be gained from the molecular 
association techniques based on linear ramps of the magnetic field 
strength across zero energy resonances, which are mainly sensitive just to 
universal scattering observables \cite{KGG03,GKGTJ03}.

Operated in the close vicinity of the zero energy resonance, the gas is
driven into a strongly correlated non-equilibrium state during the evolution 
period. In this regime the binary collision physics 
and many-body phenomena can not be disentangled because the wave function of 
the highest excited diatomic vibrational bound state strongly overlaps with 
the mean distance between the atoms in the dilute gas \cite{TKTGPSJKB03}. 
A theoretical description of the 
observations of Ref.~\cite{Claussen03} thus requires genuine many-body 
considerations, which fully account not only for the two-body physics but also 
for the non-linear dynamics of the atomic Bose-Einstein condensate in a 
non-perturbative 
manner. Our dynamical simulations of all experiments of Ref.~\cite{Claussen03} 
were based on the microscopic quantum dynamics approach to dilute 
Bose-Einstein condensates of Refs.~\cite{Koehler02,KGB03} and explained all 
the systematic trends that we were able to identify from the measured data. 
These trends involve an upward frequency shift of the interference fringes, 
accompanied by decoherence and an overall decay of the condensate. We have 
shown that the degree to which these phenomena 
influence the fringe pattern can be interpreted as a measure of the extent 
to which molecules in the highest excited diatomic vibrational bound state can 
be identified as a separate entity of the gas. Our simulations indicate that 
even in the physical regime close to the zero energy resonance, in which the 
pulse sequences are sufficiently long to reveal a decay of the condensate, the 
non-equilibrium dynamics is determined mainly by the 
same physical parameters of the binary interaction potential as in the 
interferometer mode. Deeply inelastic loss processes, such as spin 
relaxation, are likely to be negligible. Such processes will, however,
affect the parts of the experiment after the magnetic field pulse
sequence, involving the ballistic expansion of the gas before
imaging. On such longer time scales spin relaxation leads to the
spontaneous dissociation of the molecules into fast fragments, which explains the presence
of the component of undetected atoms \cite{Thompson04,Koehler04}.

\acknowledgments 
We are particularly grateful to Eleanor Hodby, Sarah Thompson and 
Carl Wieman for providing 
the experimental data to us, and to Servaas Kokkelmans for sharing with us the 
results of the coupled channels calculations of the binding energy of 
$^{85}$Rb$_2$. We thank Thomas Gasenzer and Paul Julienne for interesting 
discussions. This research has been supported through a E.C.~Marie Curie 
Fellowship under Contract no.~HPMF-CT-2002-02000 (K.G.) and a University 
Research Fellowship of the Royal Society (T.K.). K.B. thanks the Royal Society 
and the Wolfson Foundation.

\begin{appendix}
\section{Many-body dynamics of the Bose-Einstein condensate}
\label{app:Manybodydynamics}
In this appendix we derive the formulae for the frequency and visibility of 
the Ramsey fringes in the remnant Bose-Einstein condensate 
[cf.~Eqs.~(\ref{omegaeManyBody}) and (\ref{VisibilityManyBody})] from the 
first order microscopic quantum dynamics approach of 
Refs.~\cite{Koehler02,KGB03}. We restrict our considerations to the 
region of magnetic field strengths $B_\mathrm{evolve}$, in which the gas
is weakly interacting during the evolution period of the pulse sequence.
In this ideal limit of the Ramsey interferometry with atoms and molecules
in a Bose-Einstein condensate the phenomena associated with the non-linear
dynamics of the atomic mean field are weak. For simplicity, we perform all 
derivations in the homogeneous gas limit. 

\subsection{Hydrodynamic representation}
In the homogeneous gas limit the atomic mean field $\Psi(t)$ depends just on 
the time variable $t$. To study the interference fringes, it is convenient to 
represent $\Psi(t)$ in terms of its uniform density and phase:
\begin{align}
  \Psi(t)=\sqrt{n_\mathrm{c}(t)} e^{-i\varphi(t)}.
\end{align}
In this representation we then introduce the exponential form
\begin{align}
  n_\mathrm{c}(t)=e^{-2\gamma(t)}
\end{align}
of the density. The evolution of the exponent $\gamma(t)$ as well as the 
phase $\varphi(t)$ are determined simultaneously by the real and 
imaginary parts of the dynamic equation for the atomic mean field, i.e.~by
\begin{align}
  \dot{\gamma}(t)+i\dot{\varphi}(t)=-\frac{i}{\hbar}e^{2i\varphi(t)}
  \int_{t_0}^\infty d\tau \ 
  e^{-2[\gamma(\tau)+i\varphi(\tau)]}\frac{\partial}{\partial\tau}h(t,\tau),
  \label{NMNLSEhydro}
\end{align}
which can be derived directly from Eq.~(\ref{NMNLSE}) in the homogeneous gas 
limit. 

\subsection{Idealised pulse sequence}
For simplicity, we idealise the pulse sequence 
in Fig.~\ref{fig:pulse} by replacing the linear ramps of the first and second 
pulse by sudden changes of the magnetic field strength $B(t)$ at times 
$t_0<t_1<t_2<t_3=t_\mathrm{final}$, where $t_1-t_0$ and $t_3-t_2$ are the 
durations of the first and second pulse, respectively, and 
$t_\mathrm{evolve}=t_2-t_1$ is the evolution time. The 
two-body time evolution operator in Eq.~(\ref{factorisationU2B}) then 
factorises into the evolution operators at constant magnetic field strength:
\begin{align}
  U_\mathrm{2B}(t_3,t_0)=U_3(t_3-t_2)U_2(t_2-t_1)U_1(t_1-t_0).
  \label{U2B(t_3,t_0)}
\end{align}
The associated magnetic field strengths $B_1=B_\mathrm{min}=B_3$ and
$B_2=B_\mathrm{evolve}$ are shown in Fig.~\ref{fig:pulse}. At each magnetic 
field strength $B_i$ ($i=1,2,3$) of the pulse sequence, the two-body 
Hamiltonian is stationary, and $U_i(t-t_{i-1})$ can be expanded in terms of 
the stationary energy states:
\begin{align}
  \nonumber
  U_i(t-t_{i-1})=&|\phi_\mathrm{b}(B_i)\rangle 
  e^{-iE_\mathrm{b}(B_i)(t-t_{i-1})/\hbar}
  \langle\phi_\mathrm{b}(B_i)|\\
  &
  +\int d\mathbf{p}\ |\phi_\mathbf{p}^{(+)}(B_i)\rangle 
  e^{-i\frac{p^2}{m}(t-t_{i-1})/\hbar}
  \langle\phi_\mathbf{p}^{(+)}(B_i)|.
  \label{Uispectral}
\end{align}
Here $\phi_\mathrm{b}(B_i)$ is the the highest excited vibrational bound 
state whose binding energy we have denoted by $E_\mathrm{b}(B_i)$, and 
$\phi_\mathbf{p}^{(+)}(B_i)$ is the continuum energy state associated with 
the momentum $\mathbf{p}$ of the relative motion of a pair of atoms
\cite{Newton82}. Throughout this appendix we choose the continuum wave 
functions to behave at asymptotically large inter-atomic distances $r$ like
\begin{align}
  \phi_\mathbf{p}^{(+)}(\mathbf{r})\underset{r\to\infty}{\sim}
  (2\pi\hbar)^{-3/2}
  \left[
    e^{i\mathbf{p}\cdot\mathbf{r}/\hbar}+
    f(p,\vartheta)\frac{e^{ipr/\hbar}}{r}
    \right],
  \label{scattamp}
\end{align}
where $\cos\vartheta=\mathbf{p}\cdot\mathbf{r}/(pr)$ is the scattering angle,
and $f(p,\vartheta)$ is referred to as the scattering amplitude.  
The energy associated with $\phi_\mathbf{p}^{(+)}(B_i)$ is given by $p^2/m$.

To analyse the evolution of the phase $\varphi(t)$ as well as the 
exponent $\gamma(t)$, we shall successively consider each period of constant 
magnetic field strength $B_i$ ($i=1,2,3$). The dynamic equation 
(\ref{NMNLSEhydro}) can be integrated from the time $t_{i-1}$ 
of the previous sudden change of the magnetic field strength to the actual 
time $t$, which thus fulfils $t_{i-1}<t<t_i$. A separation of 
Eq.~(\ref{NMNLSEhydro}) into its real and imaginary parts then yields:
\begin{widetext}
  \begin{align}
    \label{coupledphasesphi}
    \varphi(t)&=\varphi(t_{i-1})+\frac{1}{\hbar}\mathrm{Re}
    \left(
    \int_{t_{i-1}}^{t} dt'\ e^{2i\varphi(t')}
    \left[
      -\int_{t_{i-1}}^{\infty} d\tau\ 
      e^{-2[\gamma(\tau)+i\varphi(\tau)]}
      \frac{\partial}{\partial \tau}h_i(t'-\tau)
      +C(t',t_{i-1})
      \right]
    \right),\\
    \gamma(t)&=\gamma(t_{i-1})-\frac{1}{\hbar}\mathrm{Im}
    \left(
    \int_{t_{i-1}}^{t} dt'\ e^{2i\varphi(t')}
    \left[
      -\int_{t_{i-1}}^{\infty}d\tau\ 
      e^{-2[\gamma(\tau)+i\varphi(\tau)]}
      \frac{\partial}{\partial \tau}h_i(t'-\tau)
      +C(t',t_{i-1})
      \right]
    \right).
    \label{coupledphasesgamma}
  \end{align}
\end{widetext}
Here $h_i(t'-\tau)=(2\pi\hbar)^3\langle 0|V(B_i)U_i(t'-\tau)|0\rangle
\theta(t'-\tau)$
is the coupling function of the dynamic equation (\ref{NMNLSEhydro})
restricted to the actual period $t_{i-1}<t'<t_i$ of constant magnetic field 
strength $B_i$, and 
\begin{align}
  C(t',t_{i-1})=&-\int_{t_0}^{t_{i-1}}d\tau\ \Psi^2(\tau)
  \frac{\partial}{\partial\tau}h(t',\tau)
  \label{binaryinitialcorrelations}
\end{align}  
accounts for two-body correlations produced before the previous sudden change 
of the magnetic field strength at time $t_{i-1}$. 

\subsection{Iterative solution of the non-linear Schr\"odinger equation}
We shall show that an approximation based on iteration of 
Eqs.~(\ref{coupledphasesphi}) and (\ref{coupledphasesgamma}) is sufficient
to recover the essential qualitative phenomena associated with the dynamics 
of the atomic mean field $\Psi(t)$. We thus consider the time 
integrals involving the coupling function $h_i(t'-\tau)$ and the correlation 
term $C(t',t_{i-1})$ as a small perturbation in comparison with the initial
phase $\varphi(t_{i-1})$ and the exponent $\gamma(t_{i-1})$. 
  
\begin{figure}[htb]
  \begin{center}
    \includegraphics[width=\columnwidth,clip]{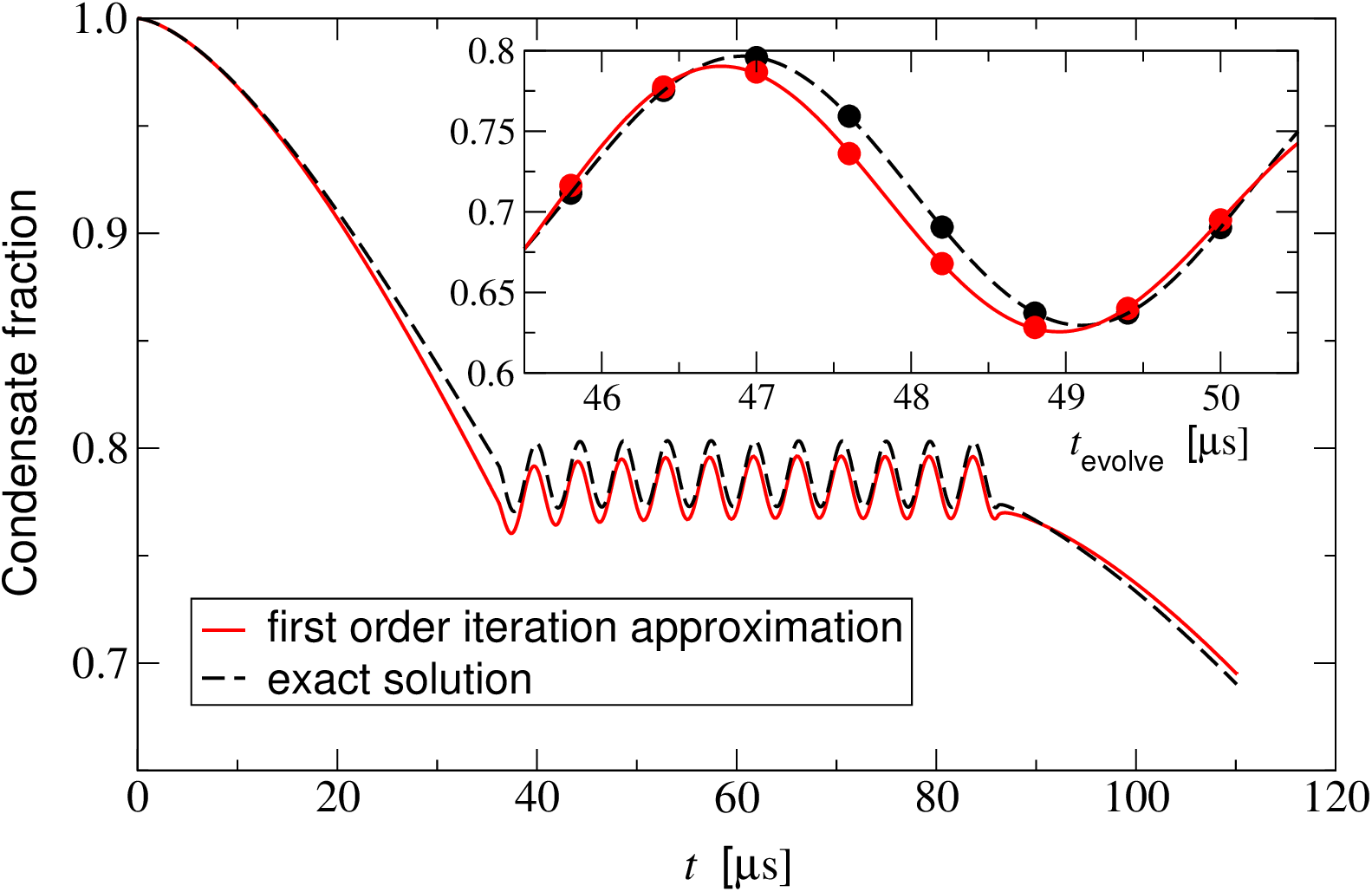}
    \caption{Comparison between the dynamics of the atomic condensate
      as obtained from the exact numerical solution (dashed curve) 
      of the non-linear Schr\"odinger equation (\ref{NMNLSEhydro}) and its 
      first order iteration approximation (solid curve) for the idealised 
      pulse sequence consisting of three periods with constant magnetic field 
      strength. The dynamics has been determined for a magnetic field strength 
      in the evolution period of $B_\mathrm{evolve}=160$ G, an evolution 
      time of $t_\mathrm{evolve}=50\, \mu$s, and an initial condensate 
      density of $n=3.16\times 10^{12}\, \mathrm{cm}^{-3}$. The inset shows a 
      comparison between the Ramsey fringes in the remnant condensate  
      as determined from the two different approaches to 
      Eq.~(\ref{NMNLSEhydro}).}
    \label{fig:analytics-check}
  \end{center}
\end{figure}

The first order perturbation 
approximation consists in replacing, on the right hand sides of 
Eqs.~(\ref{coupledphasesphi}) and (\ref{coupledphasesgamma}), the phases 
$\varphi(t')$ and $\varphi(\tau)$ as well as the density 
$n_\mathrm{c}(\tau)=\exp(-2\gamma(\tau))$ by their initial values 
$\varphi(t_{i-1})$ and $n_\mathrm{c}(t_{i-1})=\exp(-2\gamma(t_{i-1}))$, 
respectively. This procedure then shows that in the periods of constant 
magnetic field strength the approximate atomic mean field $\Psi(t)$ is subject 
to a linear time evolution
%\begin{widetext}
\begin{align}
  \Psi(t)\approx
  \exp
  \left(-\frac{i}{\hbar}\int_{t_{i-1}}^{t} dt'\
  H_\mathrm{eff}(t',t_{i-1})
  \right)\Psi(t_{i-1}).
  \label{Psifirstorder}
\end{align}
%\end{widetext}
with a complex one dimensional effective Hamiltonian
\begin{align}
  H_\mathrm{eff}(t',t_{i-1})=n_\mathrm{c}(t_{i-1})h_i(t'-t_{i-1})+
  e^{2i\varphi(t_{i-1})}C(t',t_{i-1})
  \label{Heff}
\end{align}
This gives the following expression for the density of the atomic 
Bose-Einstein condensate:
\begin{align}
  n_\mathrm{c}(t)=\exp\left(\frac{2}{\hbar}\int_{t_{i-1}}^{t} dt'\
  \mathrm{Im}\left[H_\mathrm{eff}(t',t_{i-1})\right]
  \right)n_\mathrm{c}(t_{i-1}).
  \label{ncond}
\end{align}
The dynamics of the condensate mean field is driven by the two-body 
correlations in Eq.~(\ref{binaryinitialcorrelations}), which leads to the 
explicit time dependence of $H_\mathrm{eff}(t',t_{i-1})$. Equations 
(\ref{binaryinitialcorrelations}) and (\ref{Psifirstorder}) can be used to 
produce an iterative approximate solution of the non-linear dynamic 
equation (\ref{NMNLSEhydro}) for the entire pulse sequence, just on the basis 
of the two-body coupling function. Figure \ref{fig:analytics-check} shows a 
comparison between the dynamics of the condensate predicted 
by the iterative approximation and the exact solution of 
Eq.~(\ref{NMNLSEhydro}).

%\subsection{Contributions of the initially uncorrelated part of the effective 
%  Hamiltonian}
%\label{subsec:initiallyuncorrelatedpart}
The first contribution to the sum on the right hand side of 
Eq.~(\ref{Heff}) provides the initially uncorrelated part of the effective
Hamiltonian. Its contribution to the exponent of the atomic mean field 
$\Psi(t)$ in Eq.~(\ref{Psifirstorder}) can be estimated analytically 
(cf.~Appendix \ref{app:Molecularproduction}) from 
Eq.~(\ref{Uispectral}) to be
\begin{align}
  \nonumber
  -n_\mathrm{c}(t_{i-1})\frac{i}{\hbar}\int_0^{t-t_{i-1}} d\tau\
  h_i(\tau)=&A_i(t-t_{i-1})+D_i(t-t_{i-1})\\
  &-i\mu_i(t-t_{i-1})/\hbar
  \label{Heffuncorrelated}
\end{align}
for each period of constant magnetic field strength $B_i$. Here, the leading
contribution 
\begin{align}
  \mu_i=\frac{4\pi\hbar^2}{m}n_\mathrm{c}(t_{i-1})a(B_i)
  \label{mu_i}
\end{align}
is the mean field chemical potential of the Gross-Pitaevskii equation,
\begin{align}  
  %\nonumber
  A_i(t-t_{i-1})=
  %&
  8\pi
  n_\mathrm{c}(t_{i-1})a^3(B_{i})
  %(2\pi\hbar)^3
  %\left|\langle 0|\phi_\mathrm{b}(B_i)\rangle\right|^2
  %\\
  %&\times
  \left[e^{-iE_\mathrm{b}(B_i)(t-t_{i-1})/\hbar}-1\right]
  \label{A_i}
\end{align}
is an oscillating contribution associated with the bound state spectral 
component of Eq.~(\ref{Uispectral}), and 
\begin{align}
  \nonumber
  D_i(t-t_{i-1})=&-4\pi n_\mathrm{c}(t_{i-1})a^3(B_i)
  \bigg\{
  2\sqrt{i|E_\mathrm{b}(B_i)|(t-t_{i-1})/(\pi\hbar)}
  \\
  &-
  \left[
    1-w\left(i\sqrt{i|E_\mathrm{b}(B_i)|(t-t_{i-1})/\hbar}\right)
    \right]
  \bigg\}
  %D_i(t-t_{i-1})=-4 e^{i\pi/4}
  %\left[n_\mathrm{c}(t_{i-1})a^3(B_i)\mu_i (t-t_{i-1})/\hbar\right]^{1/2}
  \label{D_i}
\end{align}
describes a diffusion in density and phase of the atomic mean field. The 
latter stems directly from the continuum part of Eq.~(\ref{Uispectral}). 
Here, we have chosen the complex square root $\sqrt{i}=\exp(i\pi/4)$ and the 
function $w(z)$ denotes the complex continuation of the error function 
\cite{Abramowitz70}:
\begin{align}
  w(z)=e^{-z^2}\mathrm{erfc}(-iz)=\frac{i}{\pi}\int\ \frac{e^{-x^2}dx}{z-x}.
  \label{wfunction}
\end{align}
The binding energies have been estimated in terms of the scattering length
$a(B_i)$ by the universal formula $E_\mathrm{b}(B_i)=-\hbar^2/[ma^2(B_i)]$.

The integral involving the initial correlations $C(t',t_0)$ vanishes 
[cf.~Eq.~(\ref{binaryinitialcorrelations})] and Eq.~(\ref{Heffuncorrelated}) 
is, therefore, sufficient to determine the evolution of $\Psi(t)$ during the 
period $t_0<t<t_1$ of the first magnetic field pulse. In our approximate 
treatment the atomic mean field at the the beginning of the evolution period 
of the idealised pulse sequence is thus given by:
\begin{align}
  \Psi(t_1)=
  e^{A_1(t_1-t_0)+D_1(t_1-t_0)}
  e^{-i\mu_1(t_1-t_0)/\hbar}\Psi(t_0).
  \label{Psi(t_1)}
\end{align}
Moreover, since we have assumed the gas to be weakly interacting during the 
evolution period, it can be verified from 
Eqs.~(\ref{Heff}), (\ref{mu_i}), (\ref{A_i}) and (\ref{D_i}) that within this 
period of time $t_1<t<t_2$ the atomic mean field in Eq.~(\ref{Psifirstorder}) 
is well approximated by the solution 
\begin{align}
  \Psi(t)=e^{-i\mu_2(t-t_1)/\hbar}\Psi(t_1)
  \label{Psievolution}
\end{align}
of the Gross-Pitaevskii equation, with the initial condensate wave function
$\Psi(t_1)$ given by Eq.~(\ref{Psi(t_1)}).

%\subsection{Contributions of the initially correlated part of the effective 
%  Hamiltonian}
%In Subsection \ref{subsec:initiallyuncorrelatedpart} we have
%We have thus successively determined the atomic mean field at all times 
%before 
%the final magnetic field pulse of the sequence, as well as the slowly varying 
%contribution of the initially uncorrelated part of the effective Hamiltonian 
%$H(t',t_2)$ to the final atomic mean field $\Psi(t_3)$. 
We shall focus in the following on the oscillatory dependence of the remnant 
condensate density $n_\mathrm{c}(t_3)$ in Eq.~(\ref{ncond}) on the evolution 
time $t_\mathrm{evolve}=t_2-t_1$, which determines the visibility and the
frequency of the Ramsey fringes. To this end, we introduce the spectral 
decomposition of the two-body time evolution 
operator $U_2(t_2-t_1)$ [cf.~Eqs.~(\ref{U2B(t_3,t_0)}) and (\ref{Uispectral})] 
in the evolution period into the, as yet undetermined, 
initially correlated part of the effective Hamiltonian 
$H_\mathrm{eff}(t',t_2)$ in Eq.~(\ref{Heff}), evaluated at times $t_2<t'<t_3$ 
during the final magnetic field pulse of the sequence. A simple calculation 
then shows that the associated contribution to the exponent of the final 
atomic mean field $\Psi(t_3)$ in Eq.~(\ref{Psifirstorder}) can be separated 
into high and low frequency components:
\begin{align}
  -\frac{i}{\hbar}\int_{t_2}^{t_3}dt\ e^{2i\varphi(t_2)}C(t,t_2)=
  A(t_\mathrm{evolve})+D(t_\mathrm{evolve}).
  \label{AhfDlf}
\end{align}
The high frequency part, i.e.
%\begin{widetext}
\begin{align}
  \nonumber
  A(t_\mathrm{evolve})=&
  (2\pi\hbar)^{3/2}
  \langle 0|U_3(t_3-t_2)|\phi_\mathrm{b}(B_2)\rangle
  e^{2i\varphi(t_2)}\\
  &\times
  \sqrt{2}\
  \Psi_\mathrm{b}(t_1+0)
  e^{i\left|E_\mathrm{b}(B_2)\right|t_\mathrm{evolve}/\hbar},
  \label{Ahf}
\end{align}
%\end{widetext}
is associated with the spectral component of $U_2(t_2-t_1)$ in the molecular
bound state $\phi_\mathrm{b}(B_2)$, and involves the molecular mean field
\cite{KGB03}
%\begin{widetext}
\begin{align}
  \nonumber
  \Psi_\mathrm{b}(t_1+0)=&-\frac{(2\pi\hbar)^{3/2}}{\sqrt{2}}
  \int_{t_0}^{t_1} d\tau \ 
  \Psi^2(\tau)\frac{\partial}{\partial\tau}
  \langle\phi_\mathrm{b}(B_2)|U_1(t_1-\tau)|0\rangle\\
  &
  +\frac{(2\pi\hbar)^{3/2}}{\sqrt{2}}
  \Psi^2(t_1)\langle\phi_\mathrm{b}(B_2)|0\rangle
  \label{Psibevolve}
\end{align}
%\end{widetext}
at the beginning of the evolution period. 
%[cf.~Eq.~(\ref{Psibgeneral})].
Furthermore, the matrix element of $U_3(t_3-t_2)$ on the right hand side of 
Eq.~(\ref{Ahf}) can be identified with the two-body probability amplitude 
%\begin{align}
%  T_\mathrm{2B}^{\#2}
%  \left(|0\rangle\leftarrow |\phi_\mathrm{b}^\mathrm{evolve}\rangle\right)
%  =\sqrt{\frac{(2\pi\hbar)^{3}}{\mathcal{V}}}
%  \langle 0|U_3(t_3-t_2)|\phi_\mathrm{b}(B_2)\rangle
%  \label{transamp2}
%\end{align}
for transitions from $|\phi_\mathrm{b}(B_2)\rangle$ to the zero energy mode 
$|\phi_0\rangle=|0\rangle\sqrt{(2\pi\hbar)^3/\mathcal{V}}$ during the second, 
final magnetic field pulse of the sequence. The probability for the 
dissociation of a molecule in the bound state 
$\phi_\mathrm{b}^\mathrm{evolve}$ into a pair of condensate atoms during 
the second magnetic field pulse is given by $P_{\mathrm{b},0}^{\#2}$
[cf.~Eq.~(\ref{transprob2})].

%\begin{align}
%  P_{\mathrm{b},0}^{\#2}=
%  \left|
%  T_\mathrm{2B}^{\#2}
%  \left(|0\rangle\leftarrow |\phi_\mathrm{b}^\mathrm{evolve}\rangle\right)
%  \right|^2.
%  \label{Pb02}
%\end{align} 

The remaining low frequency component $D(t_\mathrm{evolve})$ on 
the right hand side of Eq.~(\ref{AhfDlf}) involves an integral over relative 
momenta $\mathbf{p}$ of the amplitude associated with the burst component of 
the gas 
%[cf.~Eq.~(\ref{Psipgeneral})], 
\cite{KGB03}, which has 
evolved until the beginning of the evolution period. This diffusion part
$D(t_\mathrm{evolve})$ is thus determined mainly by the continuum 
part of $U_2(t_2-t_1)$. Its dependence on the evolution time 
$t_\mathrm{evolve}$ is comparatively weak. This allows us 
to identify the frequency as well as the visibility of the Ramsey fringes 
in the remnant condensate density from $A(t_\mathrm{evolve})$ in 
Eq.~(\ref{Ahf}), while $D(t_\mathrm{evolve})$ contributes to the average of 
the remnant condensate density with respect to the evolution time 
$t_\mathrm{evolve}$, as well as to the small decoherence of the interference 
fringes. We shall focus in the following on the determination of the 
contributions to $A(t_\mathrm{evolve})$ on the right hand side of 
Eq.~(\ref{Ahf}).

\subsection{Molecular mean field in the evolution period of the pulse 
  sequence}
To study the density and phase shift associated with the molecular mean field 
$\Psi_\mathrm{b}(t_1+0)$ in Eq.~(\ref{Ahf}), we perform a partial integration 
in Eq.~(\ref{Psibevolve}), which yields: 
\begin{align}
  \nonumber
  \Psi_\mathrm{b}(t_1+0)=&\frac{(2\pi\hbar)^{3/2}}{\sqrt{2}}
  \int_{t_0}^{t_1} d\tau \ 
  \langle\phi_\mathrm{b}(B_2)|U_1(t_1-\tau)|0\rangle
  \frac{\partial}{\partial\tau}\Psi^2(\tau)\\
  &
  +\sqrt{\frac{\mathcal{V}}{2}}\,
  T_\mathrm{2B}^{\#1}
  \left(\phi_\mathrm{b}^\mathrm{evolve}\leftarrow \phi_0\right)
  \Psi^2(t_0).
  \label{Psibevolverate}
\end{align}
The second contribution to the sum on the right hand side of 
Eq.~(\ref{Psibevolverate}) involves the probability amplitude 
%\begin{align}
%  T_\mathrm{2B}^{\#1}
%  \left(|\phi_\mathrm{b}^\mathrm{evolve}\rangle\leftarrow |0\rangle\right)
%  =\sqrt{\frac{(2\pi\hbar)^{3}}{\mathcal{V}}}
%  \langle\phi_\mathrm{b}(B_2)|U_1(t_1-t_0)|0\rangle
%  \label{transamp1}
%\end{align}
for the association of two asymptotically free zero energy ground state atoms 
into a molecular bound state 
$\phi_\mathrm{b}(B_2)=\phi_\mathrm{b}^\mathrm{evolve}$, during the first 
magnetic field pulse of the sequence [cf.~Eq.~(\ref{transamp1})]. The first 
contribution, involving the time derivative of the atomic mean field, turns 
out to be negligible. The molecular 
density $n_\mathrm{b}^\mathrm{evolve}=\left|\Psi_\mathrm{b}(t_1+0)\right|^2$
then recovers the result we found using the binary classical probability 
theory:
\begin{align}
  n_\mathrm{b}^\mathrm{evolve}\mathcal{V}=N_\mathrm{b}^\mathrm{evolve}=
  \frac{N^2}{2}\, P_{0,\mathrm{b}}^{\#1}.
\end{align}
Here $N=n_\mathrm{c}(t_0)\mathcal{V}$ and 
$N_\mathrm{b}^\mathrm{evolve}=n_\mathrm{b}^\mathrm{evolve}\mathcal{V}$
are the total number of atoms and the number of molecules in the evolution 
period, respectively, $N^2/2\approx N(N-1)/2$ is the 
number of pairs of atoms in the gas, and $P_{0,\mathrm{b}}^{\#1}$
%\begin{align}
%  P_{0,\mathrm{b}}^{\#1}=
%  \left|
%  T_\mathrm{2B}^{\#1}
%  \left(|\phi_\mathrm{b}^\mathrm{evolve}\rangle\leftarrow |0\rangle\right)
%  \right|^2
%  \label{P0b1}
%\end{align} 
is the probability for the association of a single pair of condensate atoms 
with the first magnetic field pulse of the sequence to a diatomic molecule in 
the state $\phi_\mathrm{b}^\mathrm{evolve}$ [cf.~Eq.~(\ref{transprob1})].

\subsection{Frequency of the Ramsey fringes in the remnant condensate}
\label{subsecapp:frequency}
In the first order iterative approximation to the non-linear Schr\"odinger 
equation (\ref{NMNLSEhydro}) the remnant condensate density 
$n_\mathrm{c}(t_3)=n_\mathrm{c}^\mathrm{final}(t_\mathrm{evolve})$
is determined by Eqs.~(\ref{ncond}), (\ref{Psi(t_1)}), (\ref{Psievolution}), 
and (\ref{AhfDlf}) to be
\begin{align}
  \nonumber
  n_\mathrm{c}^\mathrm{final}(t_\mathrm{evolve})=&
  %=n_\mathrm{c}(t_2)
  %\exp
  %\left(
  %\frac{2}{\hbar}\int_{t_2}^{t_3}dt\, \mathrm{Im}
  %\left[
  %  H_\mathrm{eff}(t,t_2)
  %  \right]
  %\right)
  n
  e^{2\mathrm{Re}
    \left[
      A_3(t_3-t_2)+D_3(t_3-t_2)
      \right]}
  e^{2\mathrm{Re}
    \left[
      A(t_\mathrm{evolve})+D(t_\mathrm{evolve})
      \right]}\\
  &\times
  e^{2\mathrm{Re}
    \left[
      A_1(t_1-t_0)+D_1(t_1-t_0)
      \right]},
  \label{nfinal}
\end{align}
where $n=n_\mathrm{c}(t_0)$ is the total density of the gas. Since the
diffusion term $D(t_\mathrm{evolve})$ is weakly dependent on 
$t_\mathrm{evolve}$, the oscillatory amplitude $A(t_\mathrm{evolve})$ 
determines the frequency and visibility of the Ramsey interference fringes.
We shall neglect the dependence of $D(t_\mathrm{evolve})$ on 
$t_\mathrm{evolve}$ in the following considerations and represent the two-body 
transition amplitudes of Eqs.~(\ref{transamp1}) and (\ref{transamp2}) in terms 
of their associated probabilities and phases:
\begin{align}
  T_\mathrm{2B}^{\#1}
  \left(\phi_\mathrm{b}^\mathrm{evolve}\leftarrow \phi_0\right)
  &=\left(P_{0,\mathrm{b}}^{\#1}\right)^{1/2}
  \exp\left(i\varphi_{0,\mathrm{b}}^{\#1}\right),
  \label{transamp1new}\\
  \label{transamp2new}
  T_\mathrm{2B}^{\#2}
  \left(\phi_0\leftarrow \phi_\mathrm{b}^\mathrm{evolve}\right)
  &=\left(P_{\mathrm{b},0}^{\#2}\right)^{1/2}
  \exp\left(i\varphi_{\mathrm{b},0}^{\#2}\right).
\end{align}
The oscillatory amplitude $A(t_\mathrm{evolve})$, which provides the 
only contribution to the remnant condensate density in Eq.~(\ref{nfinal})
with a significant dependence on $t_\mathrm{evolve}$, is then determined, in 
accordance with Eq.~(\ref{Ahf}), by:
\begin{align}
  A(t_\mathrm{evolve})=n\mathcal{V}
  \left(P_{\mathrm{b},0}^{\#2}P_{0,\mathrm{b}}^{\#1}\right)^{1/2}
  e^{
  %\exp
  %\left(
  i
  \left(
    \omega_\mathrm{e} t_\mathrm{evolve}+\Delta\varphi-\pi/2
    \right)}
  %\right)
  .
  \label{AtevolveMBfinal}
\end{align}
Here the total phase shift 
\begin{align}
  \Delta\varphi=\pi/2+\varphi_{0,\mathrm{b}}^{\#1}+
  \varphi_{\mathrm{b},0}^{\#2}+2\left[\varphi(t_1)-\varphi(t_0)\right],
\end{align}
consists of the sum of the phase shifts associated with the two-body 
amplitudes in Eqs.~(\ref{transamp2new}) and (\ref{transamp1new}), and
twice the phase shift of the atomic mean field accumulated during the first 
magnetic field pulse of the sequence. This can be written as
\begin{align}
  \nonumber
  2\left[\varphi(t_1)-\varphi(t_0)\right]=&
  2\mu_1(t_1-t_0)/\hbar\\
  &-2\mathrm{Im}
  \left[
    A_1(t_1-t_0)+D_1(t_1-t_0)
    \right].
\end{align}
We note that the amplitude $A(t_\mathrm{evolve})$ on the right hand side of  
Eq.~(\ref{AtevolveMBfinal}) depends just on the density of the gas rather 
than on its volume $\mathcal{V}$ because the product 
$\mathcal{V}\left(P_{\mathrm{b},0}^{\#2}P_{0,\mathrm{b}}^{\#1}\right)^{1/2}$  
is independent of $\mathcal{V}$.

The angular frequency of the fringes is determined by 
\begin{align}
  \omega_\mathrm{e}=\left|E_\mathrm{b}^\mathrm{evolve}\right|/\hbar+
  2\left[\varphi(t_2)-\varphi(t_1)\right]/(t_2-t_1),
\end{align}
where $E_\mathrm{b}^\mathrm{evolve}$ is the binding energy of the highest 
excited diatomic vibrational molecular bound state, and 
$\varphi(t_2)-\varphi(t_1)$ is the phase shift of the atomic mean field 
accumulated during the evolution period. The 
assumption of weak interactions between the atoms during the evolution 
period implies that the frequency shift due to the 
atomic mean field can be estimated in terms of the mean field chemical 
potential, i.e.~by
\begin{align}
  2\left[\varphi(t_2)-\varphi(t_1)\right]/(t_2-t_1)\approx 2\mu_2/\hbar=
  2\mu_\mathrm{evolve}/\hbar.
\end{align}
This estimate presupposes that the contributions of Eqs.~(\ref{A_i}) and 
(\ref{D_i}) to the phase shift have decayed during the evolution period. 
This stationary limit, however, is never quite reached in the experiments 
\cite{Claussen03}.

\subsection{Visibility of the Ramsey fringes in the remnant condensate}
Neglecting the dependence of the diffusion term $D(t_\mathrm{evolve})$ 
on the evolution time $t_\mathrm{evolve}$ in Eq.~(\ref{nfinal}), the 
remnant condensate density is an oscillatory function of $t_\mathrm{evolve}$
with maxima and minima at evolution times $t_\mathrm{evolve}^\mathrm{max}$ 
and $t_\mathrm{evolve}^\mathrm{min}$, which satisfy the conditions
of Eqs.~(\ref{tevolvemax}) and (\ref{tevolvemin}),
%\begin{align}
%  \sin\left(\omega_\mathrm{e} t_\mathrm{evolve}^\mathrm{max}+
%  \Delta\varphi\right)=&1,\\
%  \sin\left(\omega_\mathrm{e} t_\mathrm{evolve}^\mathrm{min}+
%  \Delta\varphi\right)=&-1,
%\end{align}
respectively. In accordance with Eq.~(\ref{nfinal}), the visibility of the 
Ramsey fringes in the remnant condensate is then given by
%\begin{widetext}
\begin{align}
  V_\mathrm{Ramsey}
  %=\frac{n_\mathrm{c}^\mathrm{final}(t_\mathrm{evolve}^\mathrm{max})-
  %  n_\mathrm{c}^\mathrm{final}(t_\mathrm{evolve}^\mathrm{min})}
  %{n_\mathrm{c}^\mathrm{final}(t_\mathrm{evolve}^\mathrm{max})+
  %  n_\mathrm{c}^\mathrm{final}(t_\mathrm{evolve}^\mathrm{min})}
  &=\frac{1-\exp
    \left(
    -2\mathrm{Re}
    \left[
      A(t_\mathrm{evolve}^\mathrm{max})- A(t_\mathrm{evolve}^\mathrm{min})
      \right]
    \right)}
  {1+\exp
    \left(
    -2\mathrm{Re}
    \left[
      A(t_\mathrm{evolve}^\mathrm{max})- A(t_\mathrm{evolve}^\mathrm{min})
      \right]
    \right)},
  %\\
  %&=\frac{1-\exp
  %  \left(
  %  -4n\mathcal{V}
  %  \left[
  %    P_{\mathrm{b},0}^{\#2}P_{0,\mathrm{b}}^{\#1}
  %    \right]^{1/2}
  %  \right)}
  %{1+\exp
  %  \left(
  %  -4n\mathcal{V}
  %  \left[
  %    P_{\mathrm{b},0}^{\#2}P_{0,\mathrm{b}}^{\#1}
  %    \right]^{1/2}
  %  \right)}.
  %\label{VisibilityManyBody}
\end{align}
%\end{widetext}
which leads directly to Eq.~(\ref{VisibilityManyBody}).

\section{Two-body transition amplitudes}
\label{app:Molecularproduction}
In this appendix we give an analytic treatment of the two-body transition 
amplitudes for molecular association and dissociation as well as for
elastic scattering at constant magnetic field strength. These amplitudes can 
be used to provide estimates of the fringe visibility and the coupling 
function $h(t,\tau)$ of the dynamic equation (\ref{NMNLSE}) for the atomic 
mean field [cf.~Eqs.~(\ref{VisibilityManyBody}) and (\ref{Heffuncorrelated}), 
respectively]. The estimates apply to an idealised pulse sequence 
consisting of periods of constant magnetic field strength with sudden changes 
in between.

\subsection{Amplitude for elastic collisions}
We first consider elastic scattering because the derivation is more 
concise than for molecular association or dissociation, while 
the basics issues are the same. In accordance with 
Eq.~(\ref{elasticampcontinuum}), the elastic transition amplitude is given by 
\begin{align}
  \nonumber
  T_\mathrm{2B}(\phi_0\leftarrow\phi_0)=&
  \langle\phi_0|U_i(t_i-t_{i-1})|\phi_0\rangle\\
  \approx &1-
  \frac{i}{\hbar\mathcal{V}}\int_0^{t_i-t_{i-1}}d\tau \ h_i(\tau).
  \label{Telastichump}
\end{align}
This result for a constant magnetic field strength 
$B_i$ ($i=1,2,3$) of an idealised pulse sequence is well approximated in 
terms of a coupling function $h_i(\tau)$ [cf.~Eq.~(\ref{couplingfunction})]
in the continuum limit. The derivation of Eq.~(\ref{Telastichump}) takes 
advantage of the general formula
\begin{align}
  \nonumber
  U_i(t_i-t_{i-1})=&U_0(t_i-t_{i-1})\\
  &+\frac{1}{i\hbar}\int_{t_{i-1}}^{t_i}dt \ 
  U_0(t_i-t)V(B_i)U_i(t-t_{i-1}),
  \label{decompositionUi}
\end{align}
where $U_0(t_i-t_{i-1})$ is the free time evolution operator. The latter can 
be expressed in terms of its spectral decomposition, i.e.
\begin{align}
  U_0(t_i-t_{i-1})=\int d\mathbf{p}\
  |\mathbf{p}\rangle e^{-i\frac{p^2}{m}(t_i-t_{i-1})/\hbar}
  \langle\mathbf{p}|.
  \label{U0spectral}
\end{align}
Here $\langle\mathbf{r}|\mathbf{p}\rangle=
\exp(i\mathbf{p}\cdot\mathbf{r}/\hbar)/(2\pi\hbar)^{3/2}$ is a plane wave 
associated with the relative momentum $\mathbf{p}$ and with the spatial
relative coordinates $\mathbf{r}$ of an atom pair. 

Since the magnetic field strength is assumed to be constant in time, the 
time evolution operator $U_i(t-t_{i-1})$ on the right hand side of 
Eq.~(\ref{decompositionUi}) is given by the spectral representation of
Eq.~(\ref{Uispectral}), which determines the coupling function to be:
\begin{align}
  \nonumber
  h_i(\tau)
  %=&(2\pi\hbar)^3\langle 0|V(B_i)U_i(\tau)|0\rangle\\
  %\nonumber
  =(2\pi\hbar)^3&
  \Bigg[\int d\mathbf{p}\ \langle 0|V(B_i)|\phi_\mathbf{p}^{(+)}(B_i)\rangle
  e^{-i\frac{p^2}{m}\tau/\hbar}\langle\phi_\mathbf{p}^{(+)}(B_i)|0\rangle\\  
  &+\langle 0|V(B_i)|\phi_\mathrm{b}(B_i)\rangle 
  e^{-iE_\mathrm{b}(B_i)\tau/\hbar}\langle\phi_\mathrm{b}(B_i)|0\rangle
  \Bigg].
  \label{h_i}
\end{align}
Here we have substituted $t-t_{i-1}=\tau$, which corresponds to the 
integration variable on the right hand side of Eq.~(\ref{Telastichump}).
We use the Lippmann-Schwinger equation \cite{Newton82}
\begin{align}
  \nonumber
  |\phi_\mathbf{p}^{(+)}(B_i)\rangle=&|\mathbf{p}\rangle+
  G_0(p^2/m+i0)V(B_i)|\phi_\mathbf{p}^{(+)}(B_i)\rangle\\
  =&|\mathbf{p}\rangle+
  G_0(p^2/m+i0)T(p^2/m+i0)|\mathbf{p}\rangle
  \label{LSE}
\end{align}
to relate the continuum energy states of the relative motion of an atom pair 
$\phi_\mathbf{p}^{(+)}(B_i)$ to the energy dependent $T$ matrix. Here 
\begin{align}
  G_0(z)=
  \left(
  z+\hbar^2\nabla^2/m
  \right)^{-1}
  =\int d\mathbf{p}\ |\mathbf{p}\rangle\frac{1}{z-p^2/m}\langle\mathbf{p}|
\end{align}
is the interaction free Green's function and ``$z=p^2/m+i0$'' indicates that
the energy argument approaches the kinetic energy $p^2/m$ from the upper half
of the complex plane. The choice of this particular set of continuum energy 
states is consistent with the asymptotic behaviour of their wave functions, 
at large distances $r$ between the atoms, as described by Eq.~(\ref{scattamp}).

The plane wave matrix elements of the energy dependent $T$ matrix can be 
expanded in terms of the parameter $pa(B_i)/\hbar$, which to first order 
accuracy leads to the universal self consistent approximation
\begin{align}
  (2\pi\hbar)^3\langle\mathbf{p}_\mathrm{out}|T(p^2/m+i0)
  |\mathbf{p}_\mathrm{in}\rangle
  \approx
  \frac{4\pi\hbar^2a(B_i)/m}{1+ipa(B_i)/\hbar}.
  \label{Tmatrixzerorange}
\end{align}
We note that the incoming and outgoing momenta $\mathbf{p}_\mathrm{in}$ and 
$\mathbf{p}_\mathrm{out}$, respectively, do not affect the 
approximate result even when their associated energies $p_\mathrm{in}^2/m$ and 
$p_\mathrm{out}^2/m$ do not match the energy argument $p^2/m$ of the $T$ 
matrix. Equation (\ref{Tmatrixzerorange}) is usually referred to as the 
contact potential approximation. It reflects, however, the universal 
low energy properties common to all $T$ matrices, independent of the 
particular shape of the short range potential. 

Inserting Eqs.~(\ref{LSE}) and (\ref{Tmatrixzerorange}) into Eq.~(\ref{h_i}), 
the right hand side of Eq.~(\ref{Telastichump}) can be represented in the 
form:
\begin{align}
  \nonumber
  T_\mathrm{2B}(\phi_0\leftarrow\phi_0)=&1
  +A_i(t_i-t_{i-1})+D_i(t_i-t_{i-1})\\
  &-i\mu_i(t_i-t_{i-1})/\hbar.
  \label{Telasticdecomp}
\end{align}
The last contribution on the right hand side of Eq.~(\ref{Telasticdecomp})
stems from the plane wave part of the scattering wave function in 
Eq.~(\ref{LSE}) and recovers the chemical potential $\mu_i$ of 
Eq.~(\ref{mu_i}), via the general relation
\begin{align}
  4\pi\hbar^2 a(B_i)/m=
  (2\pi\hbar)^3\langle 0|V(B_i)|\phi_0^{(+)}(B_i)\rangle,
\end{align}
when the volume $\mathcal{V}$ is identified with the inverse condensate 
density (cf.~Section \ref{sec:PhysicaloriginoftheRamseyfringes}).

The remaining part of the momentum integral over continuum states on the right 
hand side of Eq.~(\ref{h_i}) involves complex Gaussian integrals and determines
the diffusion term $D_i(t_i-t_{i-1})$. Expressed in terms of the complex error 
function of Eq.~(\ref{wfunction}), the diffusion term is given by
\begin{widetext}
  \begin{align}
    \nonumber
    D_i(t_i-t_{i-1})&=-\frac{i(2\pi\hbar)^3}{\hbar\mathcal{V}}
    \int_0^{t_i-t_{i-1}}d\tau
    \int \frac{d\mathbf{p}}{p^2/m}
    \left|\langle 0|V(B_i)|\phi_\mathbf{p}^{(+)}(B_i)\rangle\right|^2
    e^{-i\frac{p^2}{m}\tau/\hbar}
    %\langle\phi_\mathbf{p}^{(+)}(B_i)|V(B_i)|0\rangle
    \\
    &=-\frac{i4\pi\hbar a(B_i)}{m\mathcal{V}}
    \int_0^{t_i-t_{i-1}}d\tau\
    w\left(i\sqrt{i|E_\mathrm{b}(B_i)|\tau/\hbar}\right)
    %e^{i|E_\mathrm{b}(B_i)|\tau/\hbar}
    %\mathrm{erfc}\left(\sqrt{i|E_\mathrm{b}(B_i)|\tau/\hbar}\right)
    \label{D_iderivation}
  \end{align}
\end{widetext}
in the contact potential approximation. Here we have used Eq.~(\ref{LSE})
as well as the universal relation 
\begin{align}
  E_\mathrm{b}(B_i)=-\hbar^2/\left[ma^2(B_i)\right]
\end{align}
between the binding energy and the scattering length in the contact potential
approximation (cf.~Fig.~\ref{fig:Eb85Rb}). The remaining time integral on the 
right hand side of Eq.~(\ref{D_iderivation}) is readily performed using 
\begin{align}
  \int_0^x dy\ w\left(i\sqrt{iy}\right)=i
  \left[
    1-2\sqrt{\frac{ix}{\pi}}-w\left(i\sqrt{ix}\right)
    \right],
\end{align}
which leads directly to Eq.~(\ref{D_i}).

The bound state contribution of Eq.~(\ref{Telasticdecomp}) is found
from Eq.~(\ref{h_i}) to be
\begin{align}
  A_i(t_i-t_{i-1})=
  \frac{(2\pi\hbar)^3|\langle 0|\phi_\mathrm{b}(B_i)\rangle|^2}{\mathcal{V}}
  \left[e^{-iE_\mathrm{b}(B_i)(t_i-t_{i-1})/\hbar}-1\right],
  \label{A_iderivation}
\end{align}
where we have used the momentum space Schr\"odinger equation
\begin{align}
  \langle\mathbf{p}|V(B_i)|\phi_\mathrm{b}(B_i)\rangle=
  \left[E_\mathrm{b}(B_i)-p^2/m\right]
  \langle\mathbf{p}|\phi_\mathrm{b}(B_i)\rangle
  \label{SEphib}
\end{align}
of the bound state wave function. Inserting the universal formula  
\begin{align}
  \phi_\mathrm{b}(r)=\frac{1}{\sqrt{2\pi a(B_i)}}\frac{e^{-r/a(B_i)}}{r}
\end{align}
for the bound state wave function in the contact potential approximation
into Eq.~(\ref{A_iderivation}) yields Eq.~(\ref{A_i}).

\begin{figure}[htb]
  \begin{center}
    \includegraphics[width=\columnwidth,clip]{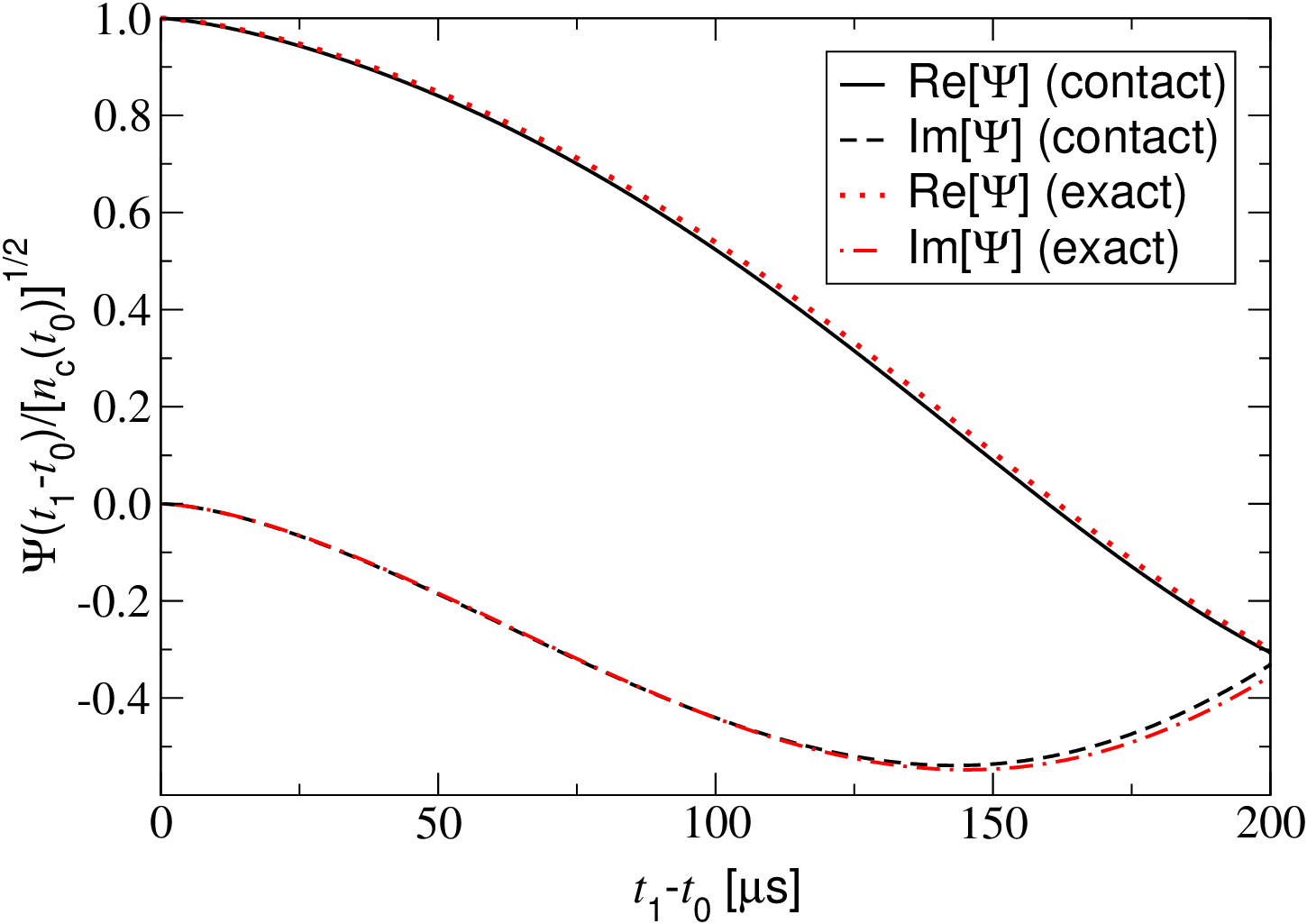}
    \caption{Comparison between the contact potential approximation and 
      exact numerical calculations of the two-body time evolution operator 
      \cite{KGB03,KGG03,GKGTJ03}
      in applications to the first order iterative approximation of the 
      atomic mean field $\Psi(t)$ [cf.~Eq.~(\ref{Psifirstorder})]. 
      The numerical calculations account for the length scales set by the 
      scattering length and by the van der Waals interaction 
      (cf.~Fig.~\ref{fig:Eb85Rb}). The initial condensate density of 
      $n_\mathrm{c}(t_0)=3.16\times 10^{12}\, \mathrm{cm}^{-3}$ is obtained 
from the low density 
      experiments of Ref.~\cite{Donley02}, while the stationary magnetic 
      field strength of $B_1=155.5$ G corresponds to the typical minimum 
      of the first pulse of the sequence in Fig.~\ref{fig:pulse}.}
    \label{fig:psiapprox}
  \end{center}
\end{figure}

We note that in applications to time independent low energy two-body collision 
phenomena the contact potential approximation is valid only in a small region 
of magnetic field strengths in the close vicinity of the zero energy 
resonance, as indicated in Fig.~\ref{fig:Eb85Rb}. In this region the 
scattering length exceeds by far all the other length scales set by the 
inter-atomic interactions. A second restriction of the contact potential 
approximation, specific to the considerations in this paper, is related to 
the fact that dynamical phenomena involve a whole range of inter-particle 
collision energies $E$. The approximation is applicable only at low 
collision energies, which depend on the magnitude of the scattering length 
through $p_\mathrm{max}a(B_i)/\hbar\ll 1$. Here 
$p_\mathrm{max}=\sqrt{m|E_\mathrm{max}|}$ denotes an estimate of the 
maximum momentum scale set by the time evolution of the observed quantity. 
Figure \ref{fig:psiapprox} compares the contact potential approximation with 
exact numerical calculations \cite{KGB03,KGG03,GKGTJ03} 
of the two-body time evolution in applications 
to the first order iterative approximation of the atomic mean field $\Psi(t)$ 
(cf.~Appendix \ref{app:Manybodydynamics}). The comparison confirms that the 
approximations leading to the contact potential approach are well 
satisfied for the elastic transition matrix element of Eq.~(\ref{Telastichump})
at the typical magnetic field strength of $B_1=155.5$ G during the first 
pulse of the sequence (cf.~Fig.~\ref{fig:pulse}).

\subsection{Molecular production and dissociation in magnetic field pulses}
In this subsection we estimate the probabilities $P_{0,\mathrm{b}}^{\#1}$ and 
$P_{\mathrm{b},0}^{\#2}$, in the contact potential approximation, for 
idealised magnetic field pulses, which just involve sudden changes of the 
magnetic field strength. The derivations provide analytic estimates 
that reveal the sensitivity of the fringe visibilities
[cf.~Section \ref{sec:PhysicaloriginoftheRamseyfringes}] to the shape of the 
magnetic field pulses. We shall consider just the amplitude 
$T_\mathrm{2B}^{\# 1}(\phi_\mathrm{b}^\mathrm{evolve}\leftarrow\phi_0)$
of the first magnetic field pulse, whose time evolution operator 
$U_{1}(t_1-t_0)=U_{\# 1}(t_1,t_0)$ is translationally invariant in time, 
under the assumption of an idealised sequence. The second amplitude 
$T_\mathrm{2B}^{\#2}(\phi_0\leftarrow \phi_\mathrm{b}^\mathrm{evolve})$
can then be obtained from considerations of time reversal. 

As the determination of 
$T_\mathrm{2B}^{\# 1}(\phi_\mathrm{b}^\mathrm{evolve}\leftarrow\phi_0)$
closely follows the considerations in the preceding section, we shall
just summarise briefly the main elements of the derivation. Taking advantage 
of the decomposition of the time evolution operator $U_{1}(t_1-t_0)$ in 
Eq.~(\ref{decompositionUi}), the transition amplitude in the continuum limit
is given by:
\begin{widetext}
  \begin{align}
    T_\mathrm{2B}^{\# 1}(\phi_\mathrm{b}^\mathrm{evolve}\leftarrow\phi_0)
    %=\langle\phi_\mathrm{b}^\mathrm{evolve}|U_{1}(t_1-t_0)|\phi_0\rangle
    \approx
    \sqrt{\frac{(2\pi\hbar)^3}{\mathcal{V}}}
    \left[
      \langle\phi_\mathrm{b}(B_2)|0\rangle+
      \frac{1}{i\hbar}\int_{0}^{t_1-t_0}d\tau\ 
      \langle\phi_\mathrm{b}(B_2)|U_0(t_1-t_0-\tau)V(B_1)U_1(\tau)|0\rangle
      \right].
    \label{T0bderive}
  \end{align}
\end{widetext}
In analogy to the derivation of the amplitude for elastic collisions, we 
replace the time evolution operators $U_{1}(\tau)$ and $U_0(t_1-t_0-\tau)$ 
by their spectral decompositions in Eqs.~(\ref{Uispectral}) and 
(\ref{U0spectral}), respectively. The energy dependent $T$ matrix elements 
that result from the continuum part of the two-body spectrum can be 
approximated in complete analogy to the preceding section. The momentum space 
bound state wave function can be determined from the Schr\"odinger equation 
(\ref{SEphib}) to be
\begin{align}
  \nonumber
  \langle\phi_\mathrm{b}(B)|\mathbf{p}\rangle
  &=\frac{\langle\phi_\mathrm{b}(B)|V(B)|\mathbf{p}\rangle}
  {E_\mathrm{b}(B)-p^2/m}\\
  &\approx
  \frac{E_\mathrm{b}(B)}
    {\pi(m|E_\mathrm{b}(B)|)^{3/4}[E_\mathrm{b}(B)-p^2/m]}
\end{align}
in the contact potential approximation. The resulting momentum and time 
integrals are similar to those in the preceding section.

In the contact potential approximation, the transition amplitude of 
Eq.~(\ref{T0bderive}) can be expressed in terms of; the scattering 
length $a(B_1)$ of the first magnetic field pulse, the ratio of scattering 
lengths
\begin{align}
  \gamma=a(B_2)/a(B_1)
\end{align}
associated with the first magnetic field pulse and the evolution period,
respectively,
in addition to the phase shifts $\delta_1=|E_\mathrm{b}(B_1)|(t_1-t_0)/\hbar$ 
and $\delta_2=|E_\mathrm{b}(B_2)|(t_1-t_0)/\hbar$. We note that just three of 
these parameters can be independently varied. The final formula for the 
transition amplitude then reads [cf.~Eq.~(\ref{wfunction}) and 
Ref.~\cite{Abramowitz70} for a discussion of the properties of the function
w(z)]:
\begin{widetext}
  \begin{align}
    \nonumber
    T_\mathrm{2B}^{\# 1}(\phi_\mathrm{b}^\mathrm{evolve}\leftarrow\phi_0)
    =\sqrt{\frac{8\pi a^3(B_1)\gamma}{\mathcal{V}}}
    \Bigg\{&
    \gamma-1+w\left(i\sqrt{i\delta_2}\right)+
    \frac{2e^{i\delta_1}}{1+\gamma}+
    \frac{2}{1-\gamma^2}
    \left[
      \gamma w\left(i\sqrt{i\delta_1}\right)-
      w\left(i\sqrt{i\delta_2}\right)
      \right]\\
    &-i\delta_2\int_0^1dx\ w\left(i\sqrt{i\delta_2(1-x)}\right)
    \left[
      \frac{1}{\sqrt{i\pi\delta_2x}}-
      w\left(i\sqrt{i\delta_2 x}\right)
      \right]
    \Bigg\}.
    \label{T2B1contact}
  \end{align}
\end{widetext}

\begin{figure}[htb]
  \begin{center}
    \includegraphics[width=\columnwidth,clip]{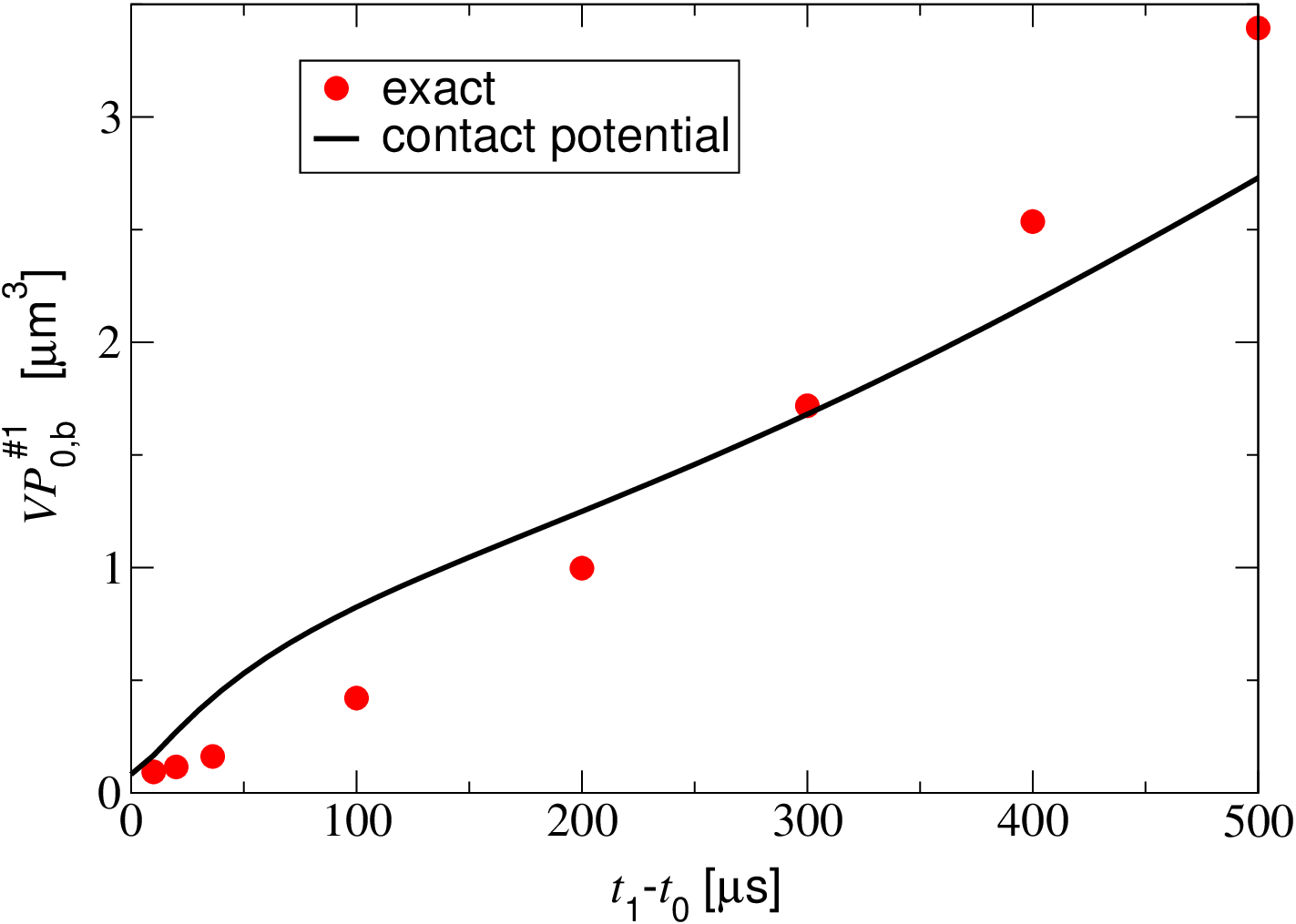}
    \caption{The molecular production efficiency 
      $P_{0,\mathrm{b}}^{\# 1}=\left|
      T_\mathrm{2B}^{\# 1}(\phi_\mathrm{b}^\mathrm{evolve}\leftarrow\phi_0)
      \right|^2$ as a function of the hold time $t_1-t_0$ of the first 
      magnetic field pulse for a minimum magnetic field strength of 
      $B_\mathrm{min}=155.5$ G (cf.~Fig.~\protect\ref{fig:pulse}). The 
      magnetic field strength in the evolution period is 
      $B_\mathrm{evolve}=156.5$ G. The figure shows exact numerical 
      calculations (filled circles) \cite{KGB03,KGG03,GKGTJ03}
      and the contact potential approximation
      obtained from Eq.~(\ref{T2B1contact}). We note that the product 
      $\mathcal{V}P_{0,\mathrm{b}}^{\# 1}$ is independent of the volume
      $\mathcal{V}$.}
    \label{fig:VP0b}
  \end{center}
\end{figure}

The application of the analytic formula of Eq.~(\ref{T2B1contact}) to the 
molecular production efficiency 
$P_{0,\mathrm{b}}^{\# 1}=\left|
T_\mathrm{2B}^{\# 1}(\phi_\mathrm{b}^\mathrm{evolve}\leftarrow\phi_0)
\right|^2$ in Fig.~\ref{fig:VP0b} illustrates that the contact potential 
approximation follows the qualitative trends of the exact numerical result. 
Equation (\ref{T2B1contact}) clearly reveals that the efficiency of molecular 
production depends on the ratio $a(B_1)/[\mathcal{V}]^{1/3}$ of the length 
scales $a(B_1)$ set by the low energy inter-atomic interactions and the mean 
inter-atomic distance, respectively, on one hand 
(cf.~Section \ref{sec:PhysicaloriginoftheRamseyfringes} for a discussion of
the physical meaning of the volume $\mathcal{V}$ in applications to dilute 
gases), and on the ratio $\gamma=a(B_2)/a(B_1)$ on the other hand. The 
scattering length $a(B_2)$ thereby provides an estimate of twice the mean 
internuclear distance \cite{Grisenti00,TKTGPSJKB03} associated with the bound 
state wave function $\phi_\mathrm{b}^\mathrm{evolve}$ in the evolution period. 
An optimal molecular production efficiency thus requires a proper balance 
between these length scales, which can differ by orders of magnitude under the 
experimental conditions of Refs.~\cite{Donley02,Claussen03}.

We also note that Eq.~(\ref{T2B1contact}) is less accurate than 
Eq.~(\ref{Telasticdecomp}) for the elastic amplitude. The reason for this is
mainly related to the different energy scales that enter the different 
amplitudes. In the case of molecular production the maximum energy scale is 
set by the binding energy $|E_\mathrm{b}(B_2)|=p_\mathrm{max}^2/m$ in the 
evolution period. The requirement 
$p_\mathrm{max}a(B_1)/\hbar\ll 1$, associated with the contact potential 
approximation, is violated in the applications. Studies beyond the scope of 
this paper show that this violation mainly affects the evolution on short 
time scales on the order of $t_1-t_0\ll \hbar/|E_\mathrm{b}(B_2)|$. Despite 
this deficit, however, the contact potential approximation provides useful
qualitative insight into the dependence of the molecular production 
efficiency on the very different length scales set by the microscopic binary 
collision physics and the mean inter-atomic distance of the gas, respectively.
\end{appendix}


\begin{thebibliography}{99}
\bibitem{footnote:Duine}
  Multiplying the field operator associated with the bare Feshbach resonance 
  level \cite{Duine03PRL} with the normalisation constant of the full 
  multichannel bound state (denoted by $\sqrt{Z(B)}$ in 
  Ref.~\cite{Duine03PRL}) does not change the state created (or annihilated) 
  by it. This state is just determined by the commutation relations of the 
  field operators \cite{GKGTJ03}. Multiplying the population of the bare 
  Feshbach resonance level with $Z(B)$ also does not in general recover the 
  measurable fraction of dressed molecules \cite{GKGTJ03} because the dressed 
  continuum states also have a component in the bare resonance level.
\bibitem{footnote}
In the experimental reference \cite{Claussen03}, the frequency shift
was attributed only to the four points closest to the zero energy
resonance. For all the other experimentally investigated values of the
magnetic field strentgh the shift was neglected and the associated
fringe frequencies used to adjust the theoretical curve of the binding
frequencies. This, in turn, determines the parameters $B_0$, $\Delta
B$ and $a_\mathrm{bg}$.
\bibitem{revisionKGB03}
  The simulations of the remnant condensate as a function of 
  $t_\mathrm{evolve}$ for an inhomogeneous gas in Fig.~11 of Ref.~\cite{KGB03} 
  were performed, by mistake, for initial conditions, which do not correspond 
  to those reported in the caption. The strong influence of inhomogeneities 
  reported in Subsection III D of Ref.~\cite{KGB03} is an artifact of this 
  mistake, and does not reflect the correct predictions of the microscopic
  quantum dynamics approach. Figure \ref{fig:85RbAMcoherence} revises 
  Fig.~11 of Ref.~\cite{KGB03}.
\bibitem{footnote:stationary}
  The existence of a stationary solution to the first order microscopic 
  quantum dynamics approach of Refs.~\cite{Koehler02,KGB03} has been 
  demonstrated in Ref.~\cite{KGG03}, and recovers the stationary 
  Gross-Pitaevskii equation in the limit of weak interactions. This refutes 
  the claim of Ref.~\cite{Duine03PRL} that the approach of 
  Refs.~\cite{Koehler02,KGB03} can not correctly incorporate equilibrium 
  properties.
\end{thebibliography}
\end{document}